\documentclass[a4paper,11pt]{article}
\usepackage[utf8]{inputenc}

\usepackage{framed}
\usepackage{mdframed}
\usepackage{psfrag}

\usepackage{fullpage}
\usepackage[T1]{fontenc}

\usepackage{amsfonts}

\usepackage[british]{babel}
\usepackage{verbatim}
\usepackage[T1]{fontenc}
\usepackage{lmodern}
\usepackage{lipsum}
\usepackage{booktabs}
\usepackage{caption}
\usepackage{cite}
\usepackage{soul,color}
\usepackage[toc,page]{appendix}
\usepackage{tikz,pgfplots}
\usetikzlibrary{decorations.pathreplacing,calc,tikzmark}

\usepackage{ifpdf}

\usepackage{amsthm, amssymb}
\usepackage[tbtags]{amsmath}
\usepackage{bm}
\usepackage{mathtools}
\usepackage{amstext}
\usepackage{braket}
\usepackage{multirow}
\usepackage[hidelinks]{hyperref}

\usepackage[normalem]{ulem}
\usepackage{xcolor}

\begin{document}
\vspace{5mm}
\vspace{0.5cm}

\def\be{\begin{eqnarray}}
\def\ee{\end{eqnarray}}
\newcommand{\eq}[1]{\begin{equation}#1\end{equation}}

\def\ba{\begin{aligned}}
\def\ea{\end{aligned}}

\def\ls{\left[}
\def\rs{\right]}
\def\lc{\left\{}
\def\rc{\right\}}

\def\p{\partial}

\newcommand{\al}[1]{\begin{align}#1\end{align}}

\def\S{\Sigma}

\def\s{\sigma}

\def\O{\Omega}

\def\a{\alpha}
\def\b{\beta}
\def\g{\gamma}

\def\ad{{\dot \alpha}}
\def\bd{{\dot \beta}}
\def\gd{{\dot \gamma}}
\newcommand{\ft}[2]{{\textstyle\frac{#1}{#2}}}
\def\ib{{\overline \imath}}
\def\jb{{\overline \jmath}}
\def\Re{\mathop{\rm Re}\nolimits}
\def\Im{\mathop{\rm Im}\nolimits}
\def\trace{\mathop{\rm Tr}\nolimits}
\def\rmi{{ i}}
\def\N{\mathcal{N}}

\newcommand{\SU}{\mathop{\rm SU}}
\newcommand{\SO}{\mathop{\rm SO}}
\newcommand{\U}{\mathop{\rm {}U}}
\newcommand{\USp}{\mathop{\rm {}USp}}
\newcommand{\OSp}{\mathop{\rm {}OSp}}
\newcommand{\Symp}{\mathop{\rm {}Sp}}
\newcommand{\Sl}{\mathop{\rm {}S}\ell }
\newcommand{\Gl}{\mathop{\rm {}G}\ell }
\newcommand{\Spin}{\mathop{\rm {}Spin}}

\def\hc{c.c.}

\numberwithin{equation}{section}

\allowdisplaybreaks

\allowbreak

\setcounter{tocdepth}{2}

%%%%%%%%%%%%%%%

\begin{titlepage}
	\thispagestyle{empty}
	\begin{flushright}

		\hfill{LMU-ASC 09/23}
		
		\hfill{MPP-2023-26}

	\end{flushright}
\vspace{35pt}

	\begin{center}
	    { \Large\bf{ 
	 On the correspondence between  black holes, 
	 }}
	 \end{center} 
	  \begin{center}
	    { \Large\bf{ 
	 domain walls and fluxes 

	     }}

		\vspace{40pt}

		{Niccol\`o~Cribiori$^{1}$,  Alessandra~Gnecchi$^{2}$, Dieter~L\"ust$^{1,3}$ and Marco~Scalisi$^{1}$
		}

		\vspace{25pt}

		{
			$^1${\it Max-Planck-Institut f\"ur Physik (Werner-Heisenberg-Institut),\\ F\"ohringer Ring 6, 80805, M\"unchen, Germany }

		\vspace{15pt}
            $^2${\it INFN, Sezione di Padova,\\ Via Marzolo, 8, 35131 Padova, Italy}

        \vspace{15pt}
            $^3${\it Arnold-Sommerfeld-Center for Theoretical Physics,\\ Ludwig-Maximilians-Universit\"at, 80333 M\"unchen, Germany }
 		}

		\vspace{40pt}

		{ABSTRACT}
	\end{center}

	\vspace{10pt}

We revisit and extend the correspondence between black holes, domain walls and fluxes in type IIA compactifications. 
We argue that these three systems can be described by the same supergravity effective action, modulo proper identifications and adjustments. 
Then, we apply the correspondence to investigate swampland conjectures on de Sitter and anti-de Sitter vacua, as well as on the black hole entropy. 
We show that, in certain cases, swampland conjectures can be motivated from properties of black hole solutions, such as positiveness of the entropy. 
This provides a bottom-up rationale which is complementary to the usual tests in string theory. 
When asking for an agreement between the anti-de Sitter and the black hole entropy distance conjectures, we are led to an extension of the correspondence which includes geometric fluxes and the associated Kaluza-Klein monopoles domain walls. 
Finally, we point out that the anti-de Sitter distance conjecture is naturally implemented in certain asymptotically anti-de Sitter black holes as a consequence of a constraint involving black hole charges and supergravity gauge couplings.

\bigskip

\end{titlepage}

%%%%%%%%%%%%%%%

\baselineskip 5.6 mm

%%%%%%%%%%%%%%%%%%%%

\tableofcontents

%%%%%%%%%%%%%%%%%%%%%

\newpage

%%%%%%%%%%%%%%%

\section{Introduction}

Black holes represent one of the most prominent tools to investigate properties of quantum gravity. 
Recently, there have been important efforts in deriving constraints on effective field theories (EFT) possibly arising from string theory or quantum gravity more in general. 
These ideas are part of broader program, which goes under the name of Swampland Program (see e.g.~\cite{Palti:2019pca, vanBeest:2021lhn,Agmon:2022thq} for reviews), and they have put black holes under scrutiny: by requiring a consistent description of black hole dynamics in the low energy, one can derive nontrivial properties of the EFT and its breakdown \cite{Bonnefoy:2019nzv,Cribiori:2022cho,Delgado:2022dkz,Cribiori:2022nke}.
Moreover, it is a fact that various swampland conjectures can be related and motivated in the context of black holes, see e.g.~\cite{Grimm:2018ohb,Gendler:2020dfp,Hamada:2021yxy,Long:2021jlv} for recent works.

Charged black holes arise as BPS solutions within the effective supergravity description of string or M-theory compactifications. 
Their entropy is given in terms of certain combinations of the electric and magnetic black hole charges: ${\cal S}={\cal S}(q,p)$.
They are typically coupled to a set of scalar fields $\phi$, namely the moduli fields of the compactification space, whose values from infinity up to the black hole horizon are determined by the attractor equations 
\cite{Ferrara:1995ih,Ferrara:1996dd}. 
In particular, scalars at the horizon are fixed in terms of the electric and magnetic black hole charges: ${\phi_H}={\phi_H}(q,p)$.
Using these relations, one can re-express the black hole entropy in terms of the scalar fields at the horizon: ${\cal S}={\cal S}(\phi_H)$.
In case of string compactifications, it follows that the black hole entropy becomes in this way  a function of the fixed moduli fields at the horizon. 

More in general, the black hole attractor mechanism  is closely related to procedure of moduli stabilization in flux vacua of string compactifications. 
In fact, already in \cite{Curio:2000sc,Behrndt:2001qa} it was pointed out that with appropriate identifications the same (supergravity) theory can be used to describe a set of different physical systems, namely black holes and string flux vacua.  
This idea has subsequently led to several important results, such as \cite{Ooguri:2004zv,Behrndt:2001mx,Ooguri:2005vr,Kounnas:2007dd,Koerber:2008rx,Caviezel:2008ik}.
Specifically, the relation between the black hole entropy and the supergravity scalar potential can be deduced by comparing the ${\cal N}=1$ superpotential $W$ and the corresponding effective supergravity scalar potential $V_{\mathcal{N}=1}$ with the black hole central charge $Z$ and the effective black hole potential $V_{\rm BH}$ \cite{Kallosh:2005ax}.

Charged BPS black holes as well as supersymmetric flux vacua possess a microscopic description in terms of closely related (D-)brane systems. 
This was first shown in the seminal works \cite{Sen:1995in,Strominger:1996sh} and subsequently generalized, starting from \cite{Maldacena:1996gb}, to four-dimensional black holes in terms of D-branes intersecting over cycles of the internal space and particle-like in the four-dimensional spacetime.
Similarly, the related supergravity background fluxes can be described by the corresponding brane sources, which have the same internal configuration as the black hole solutions, but are co-dimension one in four dimensions. 
These are the domain walls across which the background fluxes jump from zero to their value, determined by the brane configuration.
Recently, the flux/domain wall  correspondence has been used to investigate the viability of the KKLT scenario from an holographic perspective \cite{Lust:2022lfc}. See also \cite{Apers:2022vfp} for application to other setups, especially in relation to scale separation.

Therefore, we are dealing with three microscopic systems:  particle-like branes corresponding to charged black holes in the effective action, domain walls describing transitions between different flux vacua, and also the flux vacua themselves. 
All three systems are governed by closely related central charges and (super)potentials, which determine the black hole entropy, the tension of the domain walls and the value of the supergravity flux potentials.

In this paper, we review the analogies among the three systems and formulate a dictionary relating their main physical quantities. 
Then, we exploit it to investigate swampland conjectures. 
We focus in particular on conjectures involving a scalar potential, such as de Sitter and anti-de Sitter distance conjectures.
We argue that, in some cases, swampland conjectures can be seen as direct consequences of physical properties of black holes, such as the positiveness of the entropy. 
Our findings represents thus a bottom-up rationale for the investigated swampland conjectures  \cite{Hamada:2021yxy}.

We focus on ${\cal N}=2$ supergravity with vector multiplets scalars  $(L^\Lambda, M_\Lambda)$ and abelian charges $( q_\Lambda,  p^\Lambda)$, out of which we can construct the symplectic invariant combination
\begin{equation}
\label{Wgen}
\mathcal{W} =  q_\Lambda L^\Lambda- p^\Lambda M_\Lambda  .
\end{equation} 
Modulo appropriate identifications, this function governs the dynamics of the three different microscopic systems mentioned above: 

\begin{enumerate}
\item We can interpret $( q_\Lambda, p^\Lambda)$ as charges of $U(1)$ gauge fields which are non-vanishing on the vacuum. Then, we can identify \eqref{Wgen} with the $\mathcal{N}=2$ central charge, $Z=\mathcal{W}$,  and construct a black hole potential $V_{BH}=|Z|^2+g^{i\bar\jmath}D_iZD_{\bar\jmath}\bar Z$. 
Minimization of $V_{BH}$ yields $AdS_2\times S^2$ vacua, corresponding to near-horizon geometries of extremal black holes (in particular, for BPS black holes this is equivalent to extremization of the central charge $Z$) \cite{Ferrara:1996dd}. 
Here, $\mathcal{N}=2$ supergravity is ungauged and the black holes interpolate between four-dimensional Minkowski space and $AdS_2\times S^2$. 
The non-vanishing gauge fields on the vacuum are fluxes along $S^2$, but not along the $AdS_2$ directions. 
From a higher dimensional, microscopic point of view, the abelian fluxes that enter $Z$ as electric and magnetic charges account for the entropy of these black holes. 
\item  We can interpret $( q_\Lambda, p^\Lambda)$ as parameters determining the tension $Z_{DW}=\mathcal{W}$ of a domain wall separating two regions of spacetime with a different cosmological constant. 
Then, $( q_\Lambda, p^\Lambda)$ govern the jump of the various quantities across the wall \cite{Cvetic:1992bf,Lu:1996rhb}. 
\item We can interpret $( q_\Lambda, p^\Lambda)$ as charges of U(1)$_R$ Fayet-Iliopoulos gaugings of the supergravity theory. 
The scalar potential thus generated formally reduces to an ${\cal N}=1$ form, $V_{\mathcal{N}=1}=-3|\mathcal{W}|^2+|D_i\mathcal{W}|^2$, with superpotential $\mathcal{W}$ given by \eqref{Wgen} \cite{Andrianopoli:1996cm}. 
Minimization of $V_{\mathcal{N}=1}$ yields $AdS_4$ vacua (or Minkowski, but no de Sitter \cite{Fre:2002pd}), where gauge fields are zero. 
From a microscopic point of view, the charges in $\mathcal{W}$ are fluxes of a specific compactification of string theory allowing for an unbroken $AdS_4$ factor.
\end{enumerate}

From the macroscopic/EFT point of view, we are in fact giving a prescription for moving between $AdS_4$ vacua of gauged supergravity with U(1)$_R$ gauging and $AdS_2\times S^2$ BPS black holes vacua of ungauged supergravity with non-trivial profile for the gauge fields. 
One has to identify the gauge charges appearing in the scalar potential with the black hole charges associated to the vector fields supporting the black hole solution. 
A gauged supergravity vacuum, with cosmological constant $V_{\mathcal{N}=1}(p,q)<0$, corresponds to a black hole solution in ungauged supergravity, $V_{\mathcal{N}=1}=0$, with abelian charges $p,q$. 
This is described in sections \ref{bhfdw}-\ref{effsugra}.

In general, proposing a precise dictionary for this correspondence is a challenging task, but it can hint at non-trivial results. 
First, by comparing the black hole to the supergravity potentials, we obtain in section \ref{sec_VBHtoVN1} a relation between the metric and the SU(2)-connection of quaternionic manifolds which is known to be realized in certain string compactifications \cite{DallAgata:2001brr}.
Then, because of the close relationship between $AdS_2\times S^2$ BPS black hole solutions and $AdS_4$ flux vacua, we focus on models with cubic prepotential, reviewed in section \ref{cubicF}, and argue that similar version of certain swampland conjectures, such as (anti-)de Sitter and the distance conjecture, together with their generalizations, can be formulated in both systems.  
In particular, after a brief and heuristic discussion on de Sitter conjectures in section \ref{dsconj}, we focus our attention on conjectures involving anti-de Sitter vacua and black hole entropy.
The anti-de Sitter distance conjecture (ADC) \cite{Lust:2019zwm} states that in the limit of a small cosmological constant there is an associated tower of light states.\footnote{By applying the ADC to the case of a positive cosmological constant $\Lambda \to 0$ and combining it with experimental observation, the scenario of the {\sl Dark Dimension} has recently been proposed \cite{Montero:2022prj}. 
Here, the light tower of states are the Kaluza-Klein modes of a micron-size extra fifth dimension, obeying the relation $R_5\sim (\Lambda)^{-1/4}$. 
Several aspects of this scenario have been explored further \cite{Anchordoqui:2022txe,Blumenhagen:2022zzw,Gonzalo:2022jac,Anchordoqui:2022tgp}, in particular relating the
dark energy problem to the dark matter puzzle and proposing that dark matter can be either explained by light five-dimensional primordial black holes of radius $R_{BH}\sim (\Lambda)^{-1/4}$ \cite{Anchordoqui:2022txe}
or by a tower of Kaluza-Klein gravitons \cite{Gonzalo:2022jac}. 
A correspondence between these two proposals has been put forward in \cite{Anchordoqui:2022tgp}. A relation $M\sim \Lambda^\frac18$ between the supersymmetry breaking scale and the dark energy density in the dark dimension scenario has been argued for in \cite{Anchordoqui:2023oqm}.}
A similar statement has been proposed in the form of black hole entropy distance conjecture (BHEDC) \cite{Bonnefoy:2019nzv}, asserting that in the limit of large black hole entropy, ${\cal S}\rightarrow\infty$, there should be also a related light tower of states.
In \cite{Cribiori:2022cho}, the BHEDC was further explored for extremal as well as non-extremal charged black holes, where in the limit of large entropy respectively small temperature the light tower of states was indeed identified with the Kaluza-Klein modes of the internal compact space. 

When trying to relate the ADC and the BHEDC to each other in section \ref{distanceconjectures}, we will find a puzzle. We will then show in section \ref{geomflux} how to solve it by refining the correspondence between the black potential $V_{\rm BH}$ and the flux scalar potential $V_{\mathcal{N}=1}$ to include hyper multiplets together with geometric fluxes to achieve full moduli stabilization. The domain walls dual to these additional fluxes are constructed with Kaluza-Klein monopoles in the microscopic model, as explained in section \ref{KKdw}. 
Finally, in section \ref{adcaadsbh} we point out that the ADC and the BHEDC are naturally realized in certain classes of asymptotically anti-de Sitter black holes as a consequence of a quantization condition involving the product of black hole charges and supergravity gauge couplings \cite{Romans:1991nq,Cacciatori:2009iz,DallAgata:2010ejj,Hristov:2010ri}. Conclusions and future directions are summarized in section \ref{sec:conclusions}.

\section{One macroscopic description for three microscopic systems}
\label{BHDWflux:review}

In this section, we review in some detail the prescription according to which three different microscopic systems can be described by the same macroscopic setup, modulo proper adjustments.
These system are: black holes, domain walls, and string flux vacua; they can all be captured by supergravity in a unified manner. For concreteness, we also give an explicit realization in terms of D-branes in type II string theory.

\subsection{From black holes to fluxes, via domain walls}
\label{bhfdw}

We consider four-dimensional ${\cal N}=2$ supergravity as an effective theory and we look at three different setups within this framework: BPS black holes, domain walls and flux vacua. For each of them, we present two related microscopic realizations in type II string theory compactified on a six-dimensional manifold.

\subsubsection{Black holes}

Supersymmetric black hole solutions are characterised by a central charge 
\begin{align}
Z_{BH}&= q_\Lambda L^\Lambda-p^\Lambda M_\Lambda.
\end{align}
Here, $q_\Lambda$, $p^\Lambda$ are electric and magnetic charges, while $(L^\Lambda, M_\Lambda)=e^\frac{K}{2}(X^\Lambda,F_\Lambda)$ are symplectic sections of the special K\"ahler geometry, with $\Lambda=0,\dots,n_V$, with $n_V$ being the number of vector multiplets. The scalar fields $(L^\Lambda, M_\Lambda)$ are fixed at the $AdS_2\times S^2$ black hole horizon solely in terms of the charges via the (BPS) attractor equations \cite{Ferrara:1996dd}
\begin{equation}
    \partial_i |Z_{BH}|=0, \qquad i=1,\dots,n_V.
\end{equation}
These are in fact the near horizon limit of the BPS equations of the system \cite{Ferrara:1995ih}.

Microscopically, these black holes can be realised as follows.
\begin{itemize}
\item[{\bf A.}] $D_p $-$D_{p+4}$ branes configurations, which are wrapped around $p$- and $(p+4)$-cycles of the internal manifold.
\item[{\bf B.}] $D_{p+6}$ -$D_{p+2}$ branes configurations, which are wrapped around $(p+6)$- and $(p+2)$-cycles of the internal manifold.
\end{itemize}
Both configurations are particle-like with respect to the external spacetime.

At the macroscopic level, we are considering ungauged ${\cal N}=2$ supergravity and thus these black holes are asymptotically flat. As such, they can be understood as domain walls between four-dimensional Minkowski and $AdS_2 \times S^2$ vacua.

\subsubsection{Flat domain walls}
\label{domainwalls}

For completeness, we briefly sketch the procedure for constructing supersymmetric interpolations and supersymmetric domain walls, following mainly \cite{Koerber:2008rx}. 
For both kind of setups, the starting point are ${AdS}_4$ solutions of the form:
\begin{equation}
\label{adssol}
ds^2=ds^2({AdS}_4)+ds^2(\mathcal{M}_6)~,
\end{equation}
where $\mathcal{M}_6$ is a six-dimensional manifold. 
When these geometries arise as near-horizon limits of ten-dimensional supergravity solutions with brane sources, one can construct supersymmetric interpolations between them and
\begin{equation}
\label{r31sol}
ds^2=ds^2({\mathbb{R}}^{1,3})+ds^2(\mathcal{M}_6)~,
\end{equation}
far from the sources. In this way, the brane picture provides a complementary or dual picture to the emergence of $\mathrm{AdS}_4$.

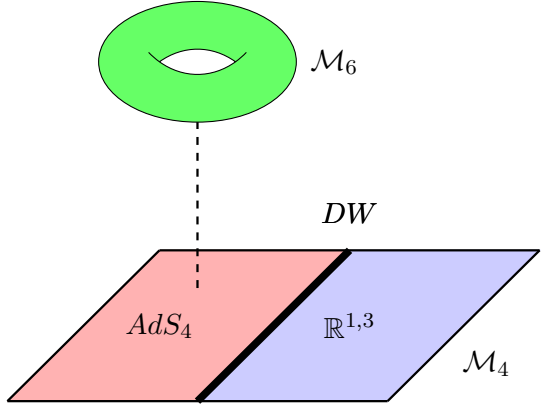
\begin{figure}[h!]
\begin{center}
\begin{tikzpicture}
\draw[fill=green!60,even odd rule] (2.5,4.5) ellipse (1.3 and .8) 
 (2.,4.5) arc(120:60:1 and 1.25) arc(-60:-120:1 and 1.25);
\draw (2.,4.5) arc(-120:-130:1 and 1.25) (3.,4.5) arc(-60:-50:1 and 1.25);
\filldraw[red!30] (0,0)--(2.5,0)--(4.5,2)--(2,2)--(0,0);
\filldraw[blue!20] (2.5,0)--(5,0)--(7,2)--(4.5,2)--(2.5,0);
\draw [line width=0.3mm] (0,0)--(5,0);
\draw [line width=0.3mm] (2,2)--(7,2);
\draw [line width=0.3mm] (0,0)--(2,2);
\draw [line width=0.3mm] (5,0)--(7,2);
\draw [line width=1mm] (2.5,0)--(4.5,2);
\draw [line width=.3mm, dashed] (2.5,1.5)--(2.5,3.7);
\node () at (2,1) {$AdS_4$};
\node () at (4.5,1) {$\mathbb{R}^{1,3}$};
\node () at (6.3,.5) {$\mathcal{M}_4$};
\node () at (4.5,2.5) {$DW$};
\node () at (4.5,2.5) {$DW$};
\node () at (4.3,4.5) {$\mathcal{M}_6$};
\end{tikzpicture}
\caption{A three-dimensional domain wall within a non-compact four dimensional space $\mathcal{M}_4$, interpolating between $AdS_4$ and $\mathbb{R}^{1,3}$. The internal manifold $\mathcal{M}_6$ is fibered over $\mathcal{M}_4$. Far away from the wall, the function $\omega(r)$ encoding the $r$-dependence of $\mathcal{M}_6$ becomes constant.}
\label{dwfig}
\end{center}
\end{figure}

Alternatively, given two solutions with different cosmological constants, such as those constructed explicitly in \cite{Kounnas:2007dd}, one can patch them together by means of a (infinitely thin) domain wall in four dimensions, as depicted in figure \ref{dwfig}.
Fluxes and the first derivative of the metric are discontinuous when crossing the three-dimensional wall patching the solutions.
A smooth metric can be obtained by departing from the infinitely-thin approximation and introducing a certain degree of thickness for the wall, namely a finite extension in the transverse directions.

The domain wall itself can be described by a metric of the form
\begin{equation}
\label{s1}
ds^2=e^{2A(r)}\left( ds^2(\mathbb{R}^{1,2}) +dr^2\right) + g_{mn}(r,y)dy^mdy^n
~,
\end{equation}
where $A(r)$ is a real function of coordinate $r$ taking values $-\infty\leq r\leq \infty$, while $y^m$ with $m=1,\dots,6$ are coordinates of $\mathcal{M}_6$. For simplicity, we assume that the internal metric $g_{mn}(r,y)$ of $\mathcal{M}_6$ is factorized as
\begin{equation}
\label{metricscale}
g_{mn}(r,y) = \omega^2(r) g_{mn}(y)
~,
\end{equation}
for some $r$-dependent function $\omega$. The metric $g_{mn}(y)$ can be for example that of a six-dimensional Calabi-Yau space. Far away from either of the sides of the wall, $r\to {\pm\infty}$, the function $\omega(r)$ behaves as
\begin{equation}
\label{dwas}
\omega(r)\to \mathrm{const} \qquad \text{for}\qquad r\to {\pm\infty}.
\end{equation}
This property distinguishes the domain wall solution from the interpolation. 

So far we have described the flow of the spacetime geometry, i.e.~the change of the metric along the transverse direction to the domain wall. 
However, in supergravity there are additional scalar fields that also flow across the wall. 
Their starting values on the Minkowski side are arbitrary constants, whereas their values on the $AdS_4$ side of the domain wall can be determined by the domain wall attractor equations. 
As explained in \cite{Curio:2000sc,Behrndt:2001qa}, these are closely related to the attractor equations of the supersymmetric black holes discussed above, upon the identification $Z_{DW}=Z_{BH}$. Namely the scalar fields on the $AdS_4$ side can be determined by the extremization of the central charge respectively the tension of the supersymmetric domain wall:
\begin{align}
Z_{DW}&= q_\Lambda L^\Lambda-p^\Lambda M_\Lambda,
\end{align}
where $q_\Lambda$, $p^\Lambda$ are now parameters determining the domain wall tension. 

To stabilize several moduli on the $AdS_4$ side of the wall, one needs to consider domain walls consisting of superpositions of different branes giving the needed amount of charges and which are embedded into the compact space in a suitable way. 
Here, we will consider type II D-branes, which will stabilize a subset of the (Calabi-Yau) scalar moduli fields.
Microscopically, these setups can be realised as follows.
\begin{itemize}
\item[{\bf A.}] $D_{p+2}$ - $D_{p+6}$ branes configurations, which are wrapped around $p$- and $(p+4)$-cycles of the internal manifold.
\item[{\bf B.}] $D_{p+8}$ - $D_{p+4}$ branes configurations, which are wrapped around $(p+6)$- and $(p+2)$-cycles of the internal manifold.
\end{itemize}
Both configurations are extending along three common directions of the external spacetime, which means that all branes act as 2-branes, i.e.~as domain walls, in the four-dimensional spacetime.

\subsubsection{Flux vacua}

The supersymmetric flux vacua of interest for us are characterised by a certain flux superpotential. 
In this respect, the domain wall picture is informative, for the tension of the wall can be viewed as the origin of this effective ${\cal N}=1$ superpotential in four dimensions. 
Going along the transverse direction and crossing the domain wall, the flux sourced by a single brane jumps by one unit.
Therefore, considering domain walls made up by bound states of branes wrapped several times around specific cycles of the internal space, the associated fluxes change accordingly. 
In the four-dimensional effective theory, this results in the generation of a flux superpotential with the form
\begin{align}
\mathcal{W}_{flux}&= q_\Lambda L^\Lambda-p^\Lambda M_\Lambda.
\end{align}

In macroscopic description, $q_\Lambda$, $p^\Lambda$ correspond to the gauging parameters determining the scalar potential of gauged $\mathcal{N}=2$ supergravity. 
In the microscopic brane picture, $q_\Lambda$ and $p^\Lambda$ are the flux numbers of the corresponding form fields, which can be realised as follows:
\begin{itemize}
\item[{\bf A.}] $(6-p)$- or dual $(2-p)$-form RR fluxes, filling  $(6-p)$- or $(2-p)$-cycles of the internal manifold.
\item[{\bf B.}] $p$- or dual $(4-p)$-form RR fluxes, filling  $p$- or $(4-p)$-cycles of the internal manifold.
\end{itemize}

Explicitly, for type IIA compactifications on Calabi-Yau orientifolds the superpotential generated by RR 6-, 4-, 2-, 0-form fluxes is \cite{Grimm:2004ua}
\begin{equation}
\begin{aligned}
W_{flux}(z) &= q_0 +q_i z^i +\frac12 C_{ijk}p^i z^j z^k - \frac{p^0}{6}C_{ijk}z^iz^jz^k,
\label{IIAF} 
\end{aligned}
\end{equation}
where $z^i$, with $i=1,\dots, h^{1,1}_-$ are K\"ahler moduli and $C_{ijk}$ the triple intersection numbers of the Calabi-Yau.\footnote{The 0-form flux corresponds to the Romans mass parameter $p^0$ in massive IIA supergravity \cite{Romans:1985tz}. Furthermore, in our conventions $X^0= 1$, $z^i=X^i/X^0$ and the $\mathcal{N}=2$ prepotential reads $F(z) = -\frac16 C_{ijk}z^iz^jz^k$.} Here and in the following we use the notation $\mathcal{W} = e^{\frac K2} W$, thus $W$ is holomorphic while $\mathcal{W}$ is covariantly holomorphic.

Notice that with this superpotential containing solely RR fluxes, only K\"ahler moduli can be stabilized, whereas the type IIA dilaton and the complex structure moduli remain flat (we will come back to this point in section \ref{sec_VBHtoVN1}).
To stabilize them one should add other fluxes, such as geometric fluxes, or NS5-branes wrapped around calibrated 3-cycles  \cite{Derendinger:2004jn,DeWolfe:2005uu,Camara:2005dc,Kounnas:2007dd}.
These additional fluxes or branes have a strong back-reaction on the internal geometry such that in general this will not be Calabi-Yau anymore, but a space with certain non-vanishing torsion classes. 
Besides, conditions from tadpole cancellation have to be obeyed.
A corresponding T-dual situation holds in type IIB compactifications with RR 3-form fluxes. Here, only complex structure moduli can be fixed and the strict Calabi-Yau geometry is not a solution of
the equations of motion \cite{Giddings:2001yu}. 
In the following, we will concentrate on the K\"ahler moduli in type IIA, namely vector multiplets scalars, since these are the fields which are determined by the ${\cal N}=2$ black hole attractor mechanism in our setups.

\subsection{An explicit realisation}

In the specific case of type IIA on a $T^2\times T^2 \times T^2$ (orientifold) with internal coordinates $y^m$, $m=1,...,6$, the systems outlined above can be described microscopically by the following D-brane and flux configurations. Related brane set-ups were also considered in \cite{Kounnas:2007dd}. We choose a coordinate frame in which the transverse direction is along $x^1=r$. Notice that the internal brane configurations A/B for black holes and domain walls are the same as internal RR-form flux configurations B/A. This is what we mean when referring to fluxes dual to branes.

\subsubsection{Black holes} 

The black hole configurations we consider can be constructed as follows.
		\begin{itemize}
			\item[\bf A.] D0-D4 system with D-branes in the following directions.
			\begin{center}
				\begin{tabular}{  l | *{4}{c} | *{6}{c}  }
				  \toprule
				  Black hole (A.) & $x^0$ & $x^1$ & $x^2$ & $x^3$ & $y^1$ & $y^2$ & $y^3$ & $y^4$ & $y^5$ & $y^6$  \\ \midrule
				  D0: & X& $\cdot$ & $\cdot$ & $\cdot$ & $\cdot$ & $\cdot$ & $\cdot$ & $\cdot$ & $\cdot$ & $\cdot$ \\ \midrule
				  D4$_1$ & X & $\cdot$ & $\cdot$ & $\cdot$ & X  & X & X  & X & $\cdot$ & $\cdot$ \\ \midrule
				  D4$_2$ & X & $\cdot$ & $\cdot$ & $\cdot$ & X & X & $\cdot$ &  $\cdot$ &  X & X \\ \midrule
				  D4$_3$  & X & $\cdot$ & $\cdot$ & $\cdot$ & $\cdot$ &  $\cdot$  & X & X &  X & X \\  
				  \bottomrule
				\end{tabular}
			  \end{center}

			\vskip0.4cm
			\noindent
			In this configuration the black hole has charges 
			  \begin{align}
				  {\rm D0}&: \qquad q_0 \ , 
				  \nonumber \\
				  {\rm D4}_i&: \qquad p^i  \ , \qquad {\rm i}=1,2,3 \ .
			  \end{align}
			\item[\bf B.] D2-D6 system with D-branes in the following directions.  
			\begin{center}
				\begin{tabular}{  l | *{4}{c} | *{6}{c}  }
				\toprule
				Black hole (B.) & $x^0$ & $x^1$ & $x^2$ & $x^3$ & $y^1$ & $y^2$ & $y^3$ & $y^4$ & $y^5$ & $y^6$  \\ \midrule
				D6:    & X & $\cdot$ & $\cdot$ & $\cdot$ & X & X & X & X & X & X\\ \midrule
				D2$_1$ & X & $\cdot$ & $\cdot$ & $\cdot$ & $\cdot$ & $\cdot$ &  $\cdot$&  $\cdot$&  X & X \\ \midrule
				D2$_2$ & X & $\cdot$ & $\cdot$ & $\cdot$ & $\cdot$ & $\cdot$  & X & X & $\cdot$ &  $\cdot$  \\ \midrule
				D2$_3$ & X & $\cdot$ & $\cdot$ & $\cdot$ & X & X & $\cdot$ &  $\cdot$ & $\cdot$ & $\cdot$  \\  
				\bottomrule
				\end{tabular}
			\end{center}
			\vskip0.4cm
			\noindent
			In this configuration the black hole has charges 
			\begin{align}
				{\rm D6}&: \qquad p^0 \ , 
				\nonumber \\
				{\rm D2}_i&: \qquad q_i  \ , \qquad i=1,2,3 \ .
			\end{align}
		\end{itemize}

\subsubsection{Domain walls}
	
 The domain walls that are related to the above black hole configurations are given as follows.
		\begin{itemize}
			\item[\bf A.] D2-D6 system with D-branes in the following internal directions.
			\begin{center}
				\begin{tabular}{  l | *{4}{c} | *{6}{c}  }
				  \toprule
				  Domain Wall (A.)& $x^0$ & $x^1$ & $x^2$ & $x^3$ & $y^1$ & $y^2$ & $y^3$ & $y^4$ & $y^5$ & $y^6$  \\ \midrule
				  D2:    & X & $\cdot$ & X & X & $\cdot$ & $\cdot$ & $\cdot$ & $\cdot$ & $\cdot$ & $\cdot$ \\ \midrule
				  D6$_1$ & X & $\cdot$ & X & X & X  & X & X  & X & $\cdot$ & $\cdot$ \\ \midrule
				  D6$_2$ & X & $\cdot$ & X & X & X & X & $\cdot$ &  $\cdot$ &  X & X \\ \midrule
				  D6$_3$ & X & $\cdot$ & X & X & $\cdot$ &  $\cdot$  & X & X &  X & X \\  
				  \bottomrule
				\end{tabular}
			  \end{center}
			\vskip0.4cm
			\noindent
			The corresponding domain wall charges are 
			\begin{align}
				{\rm D2}&: \qquad q_0 \ , 
				\nonumber \\
				{\rm D6}_i&: \qquad p^i  \ , \qquad i=1,2,3 \ .
			\end{align}
			\item[\bf B.] D8-D4 system with D-branes wrapping internal cycles in the following internal directions.
			\begin{center}
				\begin{tabular}{  l | *{4}{c} | *{6}{c}  }
				  \toprule
				  Domain Wall (B.)& $x^0$ & $x^1$ & $x^2$ & $x^3$ & $y^1$ & $y^2$ & $y^3$ & $y^4$ & $y^5$ & $y^6$  \\ \midrule
				  D8:    & X & $\cdot$ & X & X & X & X & X & X & X & X \\ \midrule
				  D4$_1$ & X & $\cdot$ & X & X & $\cdot$  & $\cdot$ & $\cdot$  & $\cdot$ & X & X \\ \midrule
				  D4$_2$ & X & $\cdot$ & X & X & $\cdot$ & $\cdot$ & X &  X & $\cdot$ & $\cdot$ \\ \midrule
				  D4$_3$ & X & $\cdot$ & X & X & X & X & $\cdot$ & $\cdot$ & $\cdot$ & $\cdot$ \\  
				  \bottomrule
				\end{tabular}
			  \end{center}
			\vskip0.4cm
			\noindent
			The corresponding domain wall charges are
			\begin{align}
				{\rm D8}&: \qquad p^0 \ , 
				\nonumber \\
				{\rm D4}_i&: \qquad q_i  \ , \qquad i=1,2,3 \ .
			\end{align}
		\end{itemize}

\subsubsection{Fluxes}
	
 Finally, the  RR fluxes corresponding to the above black hole and domain wall configurations are as follows.
	\begin{itemize}
		\item[\bf A.] 6-form/2-form fluxes in the following internal directions.
		\begin{center}
			\begin{tabular}{  l | *{4}{c} | *{6}{c}  }
			  \toprule
			 RR fluxes (A.)& $x^0$ & $x^1$ & $x^2$ & $x^3$ & $y^1$ & $y^2$ & $y^3$ & $y^4$ & $y^5$ & $y^6$  \\ \midrule
			  6-forms:    & $\cdot$ & $\cdot$ & $\cdot$ & $\cdot$ & X & X & X & X & X & X \\ \midrule
			  2-form$_1$ & $\cdot$ & $\cdot$ & $\cdot$ & $\cdot$ & $\cdot$  & $\cdot$ & $\cdot$  & $\cdot$ & X & X \\ \midrule
			  2-form$_2$ & $\cdot$ & $\cdot$ & $\cdot$ & $\cdot$ & $\cdot$ & $\cdot$ & X &  X &  $\cdot$ & $\cdot$ \\ \midrule
			  2-form$_3$ & $\cdot$ & $\cdot$ & $\cdot$ & $\cdot$ & X &  X & $\cdot$ & $\cdot$ &  $\cdot$ & $\cdot$ \\  
			  \bottomrule
			\end{tabular}
		  \end{center}
		\vskip0.4cm
			\noindent The related 
		RR flux numbers are 
		\begin{align}
			6{\textrm{-}}{\rm form}&: \qquad q_0 \ , 
			\nonumber \\
			2\textrm{-}{\rm form}_i&: \qquad p^i  \ , \qquad i=1,2,3 \ .
		\end{align}
		\item[\bf B.]  0-form/4-form fluxes in the following internal directions.
		\begin{center}
			\begin{tabular}{  l | *{4}{c} | *{6}{c}  }
			  \toprule
			  RR fluxes (B.)& $x^0$ & $x^1$ & $x^2$ & $x^3$ & $y^1$ & $y^2$ & $y^3$ & $y^4$ & $y^5$ & $y^6$  \\ \midrule
			  0-form:    & $\cdot$ & $\cdot$ & $\cdot$ & $\cdot$ & $\cdot$ & $\cdot$ & $\cdot$ & $\cdot$ & $\cdot$ & $\cdot$ \\ \midrule
			  4-form$_1$ & $\cdot$ & $\cdot$ & $\cdot$ & $\cdot$& X  & X & X  & X & $\cdot$ & $\cdot$ \\ \midrule
			  4-form$_2$ & $\cdot$& $\cdot$ & $\cdot$ & $\cdot$ & X & X & $\cdot$ &  $\cdot$ & X & X \\ \midrule
			  4-form$_3$ & $\cdot$ & $\cdot$ &$\cdot$ & $\cdot$ & $\cdot$ & $\cdot$ & X & X& X & X\\  
			  \bottomrule
			\end{tabular}
		  \end{center}
		\vskip0.4cm
			\noindent
		The non-vansihing RR flux numbers  are
		\begin{align}
			0\textrm{-form}&: \qquad p^0 \ , 
			\nonumber \\
		    4\textrm{-form}&: \qquad q_i  \ , \qquad i=1,2,3 \ .
		\end{align}
	\end{itemize}

\subsection{Effective macroscopic description of the three systems}
\label{effsugra}

We review how the three aforementioned systems are described in $\mathcal{N}=2$ supergravity in a unified manner.
For convenience, we start from the description of $AdS_4$ vacua in gauged supergravity and move then to the corresponding asymptotically flat, BPS black hole solutions governed by the same equations. 
We mainly follow the notation and conventions of \cite{Andrianopoli:1996cm}.

The bosonic action of four-dimensional gauged  $\mathcal{N}=2$ supergravity in the absence of hyper multiplets is
\begin{align}
\label{4d-FIgauged-action}
S&=\int d^4x \sqrt{-g} \left( \frac{R}{2}-g_{i\bar\jmath}\partial_\mu z^i\partial^\mu \bar z^{\bar\jmath}+\frac14{\cal I}_{\Lambda\Sigma}{\cal F}_{\mu\nu}^\Lambda {\cal F}^{\mu\nu\,\Sigma}+\frac18{\cal R}_{\Lambda\Sigma}\frac{\epsilon^{\mu\nu\rho\sigma}}{\sqrt{-g}}{\cal F}_{\mu\nu}^\Lambda {\cal F}^{\Sigma}_{\rho\sigma}-V(z,\bar z)\right) \, .
\end{align}
Here, $z^i$, with $i=1,\dots n_V$, are coordinates of a special K\"ahler manifold, the matrix $\mathcal{I}_{\Lambda\Sigma}$ is the gauge kinetic function and $\mathcal{R}_{\Lambda\Sigma}$ describes the theta-angles. The function $V(z,\bar z)$ is the scalar potential, which is proportional to the gauge coupling squared.

For the black holes we are investigating are supported by abelian charges, but the scalar fields coupled to them remain uncharged, in supergravity we have to consider constant (quaternionic) moment maps, also called Fayet-Iliopoulos terms, resulting in a U$(1)_R$ gauging. 
Since this is an R-symmetry, the gravitini are going to be charged. 
In the general case with $n_V$ scalars, there are $2n_V+2$ Fayet-Iliopoulos parameters ${\rm g}=(g_\Lambda \,,\  \tilde g^\Lambda ) $.

It is known \cite{Andrianopoli:1996cm} that the scalar potential $V(z,\bar z)$ generated by a  U$(1)_R$ gauging can be recast into an $\mathcal{N}=1$ form by defining a superpotential
\begin{equation}
\label{Wgauged}
\mathcal{W}= g_\Lambda L^\Lambda-\tilde g^\Lambda M_\Lambda,
\end{equation}
such that
\begin{align}
V(z,\bar z)=|D_i\mathcal{ W}|^2-3|\mathcal{ W}|^2 \equiv V_{\mathcal{N}=1}\ .\label{sugrapot}
\end{align}
The gauge coupling of the gravitini is given by \cite{Cribiori:2021gbf}
\begin{align}\label{g32}
g_{3/2}^2 = -\frac12 g_\Lambda {\cal I}^{\Lambda\Sigma}g_\Sigma\ 
\end{align}
(the normalization is chosen for later convenience) and it can be written in a manifestly symplectic-invariant form as 
\begin{equation}
 g_{3/2}^2 = - \frac12 {\rm g}^T \mathcal{M} {\rm g},
\end{equation}
where the matrix ${\cal M}$ is defined in equation (93) of \cite{Andrianopoli:2006ub}. This formula generalizes \eqref{g32} for it captures also magnetic gaugings.

One class of configurations we are interested in are maximally supersymmetric $AdS_4$ vacua. They are found by setting to zero the gauge fields,  ${\cal F}^\Lambda= 0$, and solving the equations
\begin{align}\label{AdS4-extr-cond}
D_i \mathcal{ W}=0\ , \qquad \Leftrightarrow \qquad \partial_i|\mathcal{ W}|=0 \ ,\qquad \text{with $\mathcal{W} \neq 0$}.
\end{align}
These are algebraic equations in the moduli allowing us to express $z^i=z^i_*(g,\tilde g)$ and thus fix the $AdS_4$ cosmological constant
\begin{align}
\Lambda =  V(z_*,\bar z_*) = -3 |\mathcal{W}(z_*,\bar z_*)|^2.
\end{align}
The ${\cal N}=2$ action \eqref{4d-FIgauged-action} admits a truncation to ${\cal N}=1$ supergravity with the same $AdS_4$ vacuum \cite{Andrianopoli:2001gm}. In the truncation, one gravitino is lost while the other has mass $m_{3/2}^2=\mathcal{W}\bar{\mathcal{W}}$; the $\mathcal{N}=1$ scalar potential is of the same form as \eqref{sugrapot}.

We now turn off the gauging, meaning that we set $g_\Lambda=\tilde g^{\Lambda}=0$ in \eqref{4d-FIgauged-action}. Thus, the scalar potential vanishes and the gravitini are uncharged. We can still look for vacua of this theory. 
If the gauge fields are vanishing, ${\cal F}^\Lambda = 0$, then the vacuum is Minkowski space $\mathbb{R}^{1,3}$ with maximal supersymmetry and the scalar fields are exact moduli.
If the gauge fields are non-vanishing, then a non-trivial vacuum exists. 
This is the well-known $AdS_2\times S^2$ geometry, supported by electric and magnetic fluxes and coupled to the scalar fields, that appear also as near horizon geometry of spherically symmetric, asymptotically flat BPS black holes. At the horizon, the value of the scalar fields is fixed as $z^i=z^i_*(q,p)$ by the attractor equations
\begin{align}
D_i { Z}=0\ , \qquad \Leftrightarrow \qquad \partial_i|Z|=0 \ ,\qquad \text{with $Z \neq 0$},
\end{align}
where 
\begin{equation}
Z=q_\Lambda L^\Lambda-p^\Lambda M_\Lambda
\end{equation}
is the central charge of the $\mathcal{N}=2$ algebra. Formally, it corresponds to the superpotential \eqref{Wgauged} of the gauged supergravity theory upon substituting ${\rm q}\to {\rm g}$, where ${\rm q}=(q_\Lambda, p^\Lambda)$ and ${\rm g}=(g_\Lambda, \tilde g^\Lambda)$. The Beckenstein-Hawking entropy is given by
\begin{equation}
\frac{S}{\pi} = |Z(z_*,\bar z_*)|^2
\end{equation}
and it is formally the same as the gravitino mass of the gauged theory, upon replacing gauge couplings with black hole charges \cite{Kallosh:2005ax}. By looking at the one-dimensional effective action of the $AdS_2\times S^2$ vacuum, one can read off the black hole potential
\begin{align}
\label{Vbhq2}
V_{BH}=-\frac12 {q}_\Lambda \mathcal{I}^{\Lambda\Sigma}{q}_\Sigma \, ,
\end{align}
or in a manifestly symplectic-invariant form
\begin{align}
V_{BH}&=-\frac12 {\rm q}^T{\cal M} {\rm q},
\end{align}
where again ${\rm q}=(q_\Lambda, p^\Lambda)$ and the matrix ${\cal M}$ can again be found in \cite{Andrianopoli:2006ub}. Upon replacing ${\rm q}\to {\rm g}$, the black hole potential corresponds to the gravitino gauge coupling of the U(1)$_R$ gauged supergravity.
The black hole potential can be also expressed in terms of the central charge
\begin{equation}
\label{BHpotential}
V_{BH} = |D_i  Z|^2 + | Z|^2\, .
\end{equation}
While extremal black holes correspond to critical points of $V_{BH}$, namely $\partial_i V_{BH}=0$, for BPS black hole we have the stronger condition $D_iZ=0$.\footnote{Similarly, vacua of supergravity are extrema of the scalar potential, but supersymmetric ones correspond also to $D_i W=0$. Besides, one can still use the condition $\partial_iV_{BH}$ to derive interesting relations using special geometry properties. For example, in \cite{Ferrara:2006em,Andrianopoli:2006ub} the authors showed that, if the scalar manifold is a symmetric space, at the extremal non-BPS horizon one has $|D_i{ Z}|=3|{ Z}|$.} Thus, for supersymmetric systems the same quantity $Z \bar Z$ is controlling the black hole potential, the entropy, the $AdS_4$ cosmological constant and the gravitino gauge coupling.
In particular, supersymmetry fixes also the $AdS_2$ and $S^2$ radii to be the same	\begin{align}
R_{S^2}^2=R_{AdS_2}^2=|{ Z}(z_*,\bar z_*)|\ .
\end{align}

\subsection{Relation between black hole and supergravity scalar potential}
\label{sec_VBHtoVN1}

We showed that, modulo adjustments which typically amount to substituting the supergravity gauge couplings with the black hole charges, the same set of equations is describing two different systems (or three, when including domain walls). 
In other words, every supersymmetric $AdS_4$ solution of $\mathcal{N}=2$ U(1)$_R$ gauged supergravity is in one-to-one correspondence with an asymptotically flat, spherically symmetric BPS black hole.
Supersymmetry plays a central role in this relation. 

It is tempting to propose a dictionary between these systems. One could associate the black hole central charge to the gravitino mass 
\begin{equation}\label{prop-dic-Z}
    Z \qquad \Longleftrightarrow \qquad m_{3/2}
\end{equation}
and the black hole potential to the supergravity potential
\begin{equation}\label{prop-dic-V}
    V_{BH}  \qquad \Longleftrightarrow \qquad V_{\mathcal{N}=1}.
\end{equation}
The identification can be done at the level of $\mathcal{N}=2$ supergravity, or more directly by considering its consistent truncation to $\mathcal{N}=1$. In any case, at the present stage the correspondence seems not to be precise, especially for what concerns the scalar potentials. To understand the issue, in the following we first assume the second viewpoint and look at the problem from the $\mathcal{N}=2$ perspective only in a second step. The latter will lead us to recover a specific property of quaternionic manifolds which is realized in certain string compactifications.

First, let us notice that, as it is well known, the condition $D_i Z=0$ implies both $\partial_iV_{BH}=0$ and $\partial_iV_{\mathcal{N}=1}=0$, but the opposite is not true. Indeed, there exists non-BPS black holes and non-supersymmetric anti-de Sitter vacua. However, these vacua cannot be reached if one is asking for an extremum of both potentials at the same time, since
\begin{align}
    \partial_k V_{BH} &= g^{i\bar\jmath} D_k D_iZ \bar D_{\bar\jmath}\bar Z + 2D_kZ \bar Z \equiv 0,\\
     \partial_k V_{\mathcal{N}=1} &= g^{i\bar\jmath} D_k D_iZ \bar D_{\bar\jmath}\bar Z - 2D_kZ \bar Z \equiv 0
\end{align}
are either solved by $D_iZ=0$, corresponding to BPS black holes and supersymmetric anti-de Sitter vacua, or by $Z=0=g^{i\bar\jmath} D_kD_i Z\bar D_{\bar\jmath}\bar Z$, corresponding to non-BPS black hole and de Sitter vacua.\footnote{De Sitter critical points with vanishing gravitino mass have been conjectured to be in the swampland in \cite{Cribiori:2020use,DallAgata:2021nnr}.}

Second, the two potentials actually differ by a term proportional to $Z\bar Z$, namely
\begin{equation}
\label{VbhVsugra}
V_{BH} = V_{\mathcal{N}=1} + 4 Z\bar Z.
\end{equation}
The reason for this discrepancy can be understood as follows. 
The correspondence between black holes and flux potentials holds for the subset of scalar fields stabilized by the black hole attractor mechanism.
In our type IIA setups with D-branes and RR fluxes, these are K\"ahler moduli. 
However, in the same type IIA Calabi-Yau compactifications, the complex structure moduli and the dilaton are not stabilized by RR fluxes; indeed, we did not include them in the effective supergravity description.
Their inclusion will allow us to explain the above discrepancy. 

For concreteness, let us consider a simple model with ${\cal N}=1$ chiral multiplets $\{T^i,U^i,S \}$, with $i=1,2,3$. Here, $T^i$ are K\"ahler moduli, $U^i$ are complex structure moduli and $S$ represents the axion-dilaton. In ${\cal N}=2$ language, K\"ahler moduli  correspond to vector multiplets (denoted $z^i$ above), while $U^i$  and $S$  correspond to hyper multiplets. We can simplify the model further by taking $T^i\equiv T$ and $U^i\equiv U$ for all values of $i$; this will not affect the argument.

One way to proceed is simply to assume that the fields $U$ and $S$ are stabilized by some mechanism that we do not specified further. 
In this case, minima of the potential \eqref{sugrapot} with respect to the K\"ahler moduli will generically have negative vacuum energy, i.e.~they will be $AdS_4$ vacua. Clearly, here the issue is how to make the stabilization mechanism concrete.

A second way to proceed is by exploiting the correspondence between flux potentials and black holes and mimic in supergravity what happens in the attractor mechanism for black holes. 
Since the attractor flow stabilizes vector multiplets but not hyper multiplets, this corresponds to having in supergravity a holomorphic superpotential $W$ depending only on the K\"ahler moduli,
\begin{equation}
    W = W(T),
\end{equation}
while for the K\"ahler potential we take 
\begin{equation}
\label{KSTU}
K= -3\log(-i(T-\bar{T}))-3\log(-i(U-\bar{U}))-\log(-i(S-\bar{S}))\,,
\end{equation}
where the coefficients $3$ arise from the sum over $i=1,2,3$. The $\mathcal{N}=1$ scalar potential of this model exhibits a no-scale structure along $U$ and $S$, for the superpotential $W$ does not depend on them. Therefore, we have
\begin{align}
 g^{U\bar{U}}{D}_U W{D}_{\bar{U}}\bar{W}&=g^{U\bar{U}}K_{U}K_{\bar{U}} |W|^2= 3|W|^2\,,\\
 g^{S\bar{S}}{D}_S W{D}_{\bar{S}}\bar{W}&=g^{S\bar{S}}K_{S}K_{\bar{S}}|W|^2= |W|^2\,,
\end{align}
meaning that the fields $U$ and $S$ are effectively contributing positively to the scalar potential, with a term $4\mathcal{W}\bar{\mathcal{W}}$ (we recall that $W$ is holomorphic, while $\mathcal{W}=e^\frac{K}{2}W$ is covariantly holomorphic). 
We can then isolate this contribution
\begin{equation}
V_{{\cal N}=1}=\hat V_{{\cal N}=1}-4\mathcal{W}\bar{\mathcal{W}}. 
\end{equation}
This defines
\begin{equation}
\hat V_{{\cal N}=1}=|D_T\mathcal{W}|^2 + |\mathcal{W}|^2,
\end{equation}
which corresponds precisely to the black hole potential \eqref{BHpotential} upon identifying the central charge $Z$ with the covariantly holomorphic superpotential $\mathcal{W}$. As long as the holomorphic potential depends only on K\"ahler moduli, one can repeat the argument for generic IIA Calabi-Yau orientifolds, for one has that $K_a g^{a\bar b} K_{\bar b} = 4$, where $a,b$ run over complex structure moduli and axio-dilaton \cite{Grimm:2004ua}. 

We now look at the problem from the $\mathcal{N}=2$ perspective. On top of the R-symmetry gauging, we consider a generic abelian gauging with non-vanishing, constant quaternionic killing vectors.
The supergravity scalar potential is then \cite{Andrianopoli:1996cm}
\begin{equation}
    V_{\mathcal{N}=2} = |D_i \mathcal{W}|^2 - 3|\mathcal{W}|^2 + 4h_{uv}k^u_\Lambda k^v_\Sigma \bar L^\Lambda L^\Sigma,
\end{equation}
where $h_{uv}$ and $k^u_\Lambda$ are the quaternionic metric and killing vectors respectively. The condition for this scalar potential to match with $V_{BH}$ is
\begin{equation}
    h_{uv}k^u_\Lambda k^v_\Sigma \bar L^\Lambda L^\Sigma =  |\mathcal{W}|^2. 
\end{equation}
This is a consequence of the relation \cite{DallAgata:2001brr}
\begin{equation}
\label{kk=PP}
     h_{uv}k^u_\Lambda k^v_\Sigma = \mathcal{P}^x_\Lambda \mathcal{P}^x_\Sigma,
\end{equation}
which indeed holds for an (electric) abelian gauging generated by constant quaternionic killing vectors.\footnote{In fact, \eqref{kk=PP} follows from the condition that $k^u_\Lambda$ are constant, implying $\mathcal{P}^x_\Lambda = - k^u_\Lambda\omega^x_u$ \cite{DAuria:1990qxt}, together with the condition $\omega^x_u\omega^x_v = h_{uv}$, where $\omega^x$ is the SU(2)-connection.} Such gauging arises naturally in type II compactifications on Calabi-Yau in the presence of background fluxes, where the charged quaternionic scalar is dual to the $B_2$ field, see e.g.~\cite{Louis:2002ny} for a comprehensive discussion. 
Thus we learned that, asking for a matching between $V_{BH}$ and $V_{\mathcal{N}=2}$ leads to the relation \eqref{kk=PP}, realized in concrete string models, and the supergravity scalar potential reduces to \cite{DallAgata:2001brr}
\begin{equation}
\label{VBHN2}
    V_{\mathcal{N}=2} = \left(U^{\Lambda \Sigma} + \bar L^\Lambda L^\Sigma\right)\mathcal{P}^x_\Lambda \mathcal{P}^x_\Sigma = -\frac12 \mathcal{I}^{\Lambda \Sigma} \mathcal{P}^x_\Lambda \mathcal{P}^x_\Sigma.
\end{equation}
This can be seen as the generalisation of \eqref{Vbhq2} in which the charges become functions of the quaternionic scalars. However, the limit of constant moment maps is not continuously connected to the present situation.
This happens because the extremization over a subspace of scalars is not automatically promoted to the extremization over the entire field space.

The scalar potential \eqref{VBHN2} is positive definite and one can prove that it does not admit de Sitter vacua, for field-dependent moment maps. Due to the factorized form in which the vector multiplet scalars appear only in the matrix $\mathcal{I}^{\Lambda \Sigma}$, while the quaternionic scalars only in the moment maps $\mathcal{P}^x_\Lambda$, we have
\begin{equation}
    \partial_u V_{\mathcal{N}=2} = {\mathcal{I}^{\Lambda\Sigma}}\Omega^x_{uv}k^v_\Lambda \mathcal{P}^x_\Sigma,
\end{equation}
where we used that $-2\Omega_{uv}^x k^v_\Lambda = \partial_u \mathcal{P}^x_\Lambda + \epsilon^{xyz}\omega_u^y\mathcal{P}^z_\Lambda\equiv \nabla_u \mathcal{P}^x_\Lambda$ and the term proportional to $\epsilon^{xyz}$ vanishes for symmetry reasons. Thus, we find a critical point in the quaternionic directions either if $k^u_\Lambda=0$ or if $ \mathcal{P}^x_\Sigma=0$. The first case can at best lead to a supersymmetric critical point, while in the second case supersymmetry is broken ($k_\Lambda^u \neq 0$) but the potential is vanishing.

\section{Models with cubic prepotential}
\label{cubicF}

In this section, we look at a specific class of models and present their main features, such as the black hole potential and the entropy of BPS black holes. These are going to be used in what follow to investigate swampland conjectures.
More details on definitions and calculations can be found in appendix \ref{app_skgeom}.

Type IIA compactifications on Calabi-Yau manifolds in the large K\"ahler moduli regime can be described by $\mathcal{N}=2$ supergravity with a cubic prepotential
\begin{equation}\label{cubic-F}
F = -\frac{1}{6} C_{ijk}\frac{X^i X^j X^k}{X^0},
\end{equation}
where $C_{ijk}$ are the triple intersection numbers of the Calabi-Yau. 
We recall that $(X^\Lambda, F_\Lambda)$ are symplectic sections, while the physical scalars of the vector multiplets can be parametrized as
\begin{equation}
    z^i = \frac{X^i}{X^0}, \qquad i=1,\dots,n_V \equiv h^{1,1}.
\end{equation}
Hyper multiplets are not described by the prepotential. The classical volume of the internal manifold is given by
\begin{equation}
    \mathcal{V} = \frac16 C_{ijk} t^i t^j t^k,
\end{equation}
where we split $z^i=b^i+it^i$. One can check that this is related to the K\"ahler potential as
\begin{equation}
    K = -\log \left[i\left(\bar X^\Lambda F_\Lambda - X^\Lambda \bar F_\Lambda\right)\right] = - \log \left(8\mathcal{V} X^0\bar X^0\right).
\end{equation}
Asking for the volumes of all complex submanifolds of the Calabi-Yau to be positive defines a region of the moduli space known as K\"ahler cone. For the supergravity approximation to be reliable, one should remain within the bulk of this region.

We are interested in BPS black holes in this setup. They are supported by electric and magnetic charges $(q_\Lambda, p^\Lambda)$, entering the central charge of the supersymmetry algebra as
\begin{equation}
    Z = q_\Lambda L^\Lambda - p^\Lambda M_\Lambda.
\end{equation}
To make contact with the microscopic description, we choose to work with non-vanishing charges $-q_0\equiv q >0$ and $p^i >0$. This choice corresponds to the brane/flux setups A of section \ref{BHDWflux:review}. A similar investigation can be performed for the setups B. The central charge is explicitly
\begin{equation}
    Z = X^0 e^{\frac K2} \left(-q + \frac12 C_{ijk}p^i z^j z^k\right).
\end{equation}
The black hole potential can be calculated as shown in appendix \ref{app_skgeom}. Setting the axions to zero, $b^i=0$, it simplifies to
\begin{equation}
\label{Vbhb=0}
    V_{BH}|_{b^i=0} = \frac{q^2}{2\mathcal{V}} + 2\mathcal{V}\, {\rm p}^2,
\end{equation}
where ${\rm p}^2=p^i g_{i\bar\jmath}p^{\bar \jmath}$.
This potential is manifestly positive definite within the K\"ahler cone.

At the black hole  horizon, the scalar fields are fixed in terms of the charges by means of the attractor equations, solved by \cite{Ferrara:1996dd,Behrndt:1996jn}
\begin{equation}
    p^\Lambda = - 2 {\rm Im}\, X^\Lambda, \qquad q_\Lambda = -2{\rm Im} F_\Lambda,
\end{equation}
giving in turn
\begin{equation}
    z_*^i=it^i=ip^i\sqrt{\frac{q}{\frac16 C_{jkm}p^jp^kp^m}}, \qquad b^i=0.
\end{equation}
We can evaluate the various quantities at the horizon. The volume of the Calabi-Yau manifold is 
\begin{equation}
\mathcal{V}_* = \sqrt{\frac{q^3}{\frac16 C_{ijk}p^ip^jp^k}},    
\end{equation}
the flux potential is
\begin{equation}
   V_{\mathcal{N}=1}(z_*,\bar z_*) \equiv \Lambda= -6 \sqrt{\frac q6 C_{ijk}p^ip^jp^k} 
\end{equation}
and finally the black hole entropy is
\begin{equation}
\frac{\mathcal{S}}{\pi} = |Z(z_*,\bar z_*)|^2 =V_{BH}(z_*,\bar z_*)  = 2 \sqrt{\frac q6 C_{ijk}p^ip^jp^k}  .
\end{equation}
For these expressions to be real within the given choice of charges, we have to ask $C_{ijk}\geq 0$. This descends from the consistency condition $C_{ijk}p^ip^jp^k \geq 0$ for supergravity strings in five dimension \cite{Katz:2020ewz}.

We can also express $\Lambda$ and ${\cal S}$ in terms of the volume ${\cal V}$. As in \cite{Cribiori:2022cho}, we consider two different cases, namely (i), where we keep the electric charge $q$ fixed and vary the magnetic charge $p^i$,
and conversely  (ii), where we keep the magnetic charge $p^i$ fixed and vary the electric charge $q$:
\begin{eqnarray}\label{SVi}
\text{ case (i):} \qquad \Lambda & = & -6{\frac{q^2}{{\cal V}_*}}\, ,
\cr
\frac{{\cal S}}{\pi}& =&2{\frac{q^2}{{\cal V}_*}}\, ,
\end{eqnarray}
and 
\begin{eqnarray}\label{SVii}
\text{ case (ii):} \qquad \Lambda & = & -6\left(\frac16 C_{ijk}p^ip^jp^k\right)^\frac23{{\cal V}_*}^{1/3}\, ,
\cr
\frac{\cal S}{\pi}& = & 2 \left(\frac16 C_{ijk}p^ip^jp^k\right)^\frac23{{\cal V}_*}^{1/3}  \, .
\end{eqnarray}
We see that these two cases exhibit 
opposite behaviours:  for case (i) the cosmological constant and the entropy decrease with increasing volume, where for case (ii) these two quantities increase for growing volume.
We will discuss this further in section (\ref{distanceconjectures}).

\section{Swampland conjectures on the potentials}

In this section, we apply the correspondence between black holes, domain walls and flux potentials developed so far to test and motivate and recover swampland conjectures. We focus on conjectures involving a scalar potential, in particular de Sitter conjectures and anti-de Sitter/entropy distance conjectures. The investigation of the latter will uncover a puzzle which is then solved in the next section.

\subsection{De Sitter conjectures}
\label{dsconj}

First, let us study the extrema of the potentials $V_{BH}$ and $V_{\mathcal{N}=1}$; in what follows, we systematically set $b^i=0$. The potential $V_{BH}$ is given in \eqref{Vbhb=0}, while $V_{\mathcal{N}=1}$ is
\begin{equation}
\label{VN1b0}
    V_{\mathcal{N}=1}= -\frac12 C_{ijk}p^ip^j t^k - \frac{q}{2\mathcal{V}}C_{ijk}t^it^jp^k.
\end{equation}
Using that 
\begin{equation}
t^i\frac{\partial}{\partial t^i} \mathcal{V} = 3\mathcal{V},\qquad t^i\frac{\partial}{\partial t^i} {\rm p}^2 = -2 {\rm p}^2,
\end{equation}
following from\eqref{tdg}, we get
\begin{align}
    t^i \frac{\partial}{\partial t^i} V_{BH} &= - 3 V_{BH} + 8 \mathcal{V}\, {\rm p}^2,\\
 t^i \frac{\partial}{\partial t^i} V_{\mathcal{N}=1} &= V_{\mathcal{N}=1} + \frac{q}{\mathcal{V}}C_{ijk}t^it^jp^k,    
\end{align}
implying that the vacuum energies are respectively
\begin{align}
    V_{BH} &= \frac83\mathcal{V}\,{\rm p}^2,\\
    V_{\mathcal{N}=1} &= - \frac{q}{\mathcal{V}}C_{ijk}t^it^jp^k.
\end{align}
We see again that, in the K\"ahler cone, the vacuum energy of the black hole potential is positive definite regardless of the choice of the charges. This is indeed necessary for the entropy to be positive, even for non-BPS black holes. Instead, the sign of the vacuum energy of the $ V_{\mathcal{N}=1}$ potential is not fixed a priori. Insisting that the right hand side admits a black hole interpretation, we are led to a negative definite cosmological constant. In fact, for the potential $V_{\mathcal{N}=1}$ to have a negative definite sign at any point of the moduli space (with $b^i=0$), one has to introduce further assumptions on the supergravity side, as for example $C_{ijk}>0$ implying $t^i>0$ within the K\"ahler cone  (the first condition follows if the K\"ahler cone is simplicial), or viceversa. These assumptions are automatic in the black hole picture.

We have seen that insisting for a black hole interpretation highly constrains the presence of de Sitter critical points in the supergravity potential. On the other hand, the (refined) de Sitter conjectures postulate specific bounds on the scalar potential and its derivatives. Here, we focus on the proposal of \cite{Ooguri:2018wrx}, namely
\begin{align}\label{ineq-1}
	V' _{{\cal N}=1}> c_1 V_{{\cal N}=1} \ ,
	\qquad \text{or} \qquad
	V''_{{\cal N}=1} < - c_2 V_{{\cal N}=1} \ ,
\end{align}
with $c_{1,2}$ positive order one parameters. These relations forbid (stable) de Sitter vacua and it would be interesting to derive them from black hole arguments. However, this seems not straightforward to us. For example, we could turn the logic around and try to see if some function in supergravity satisfies these conjectures as a consequence of properties of the black hole potential. 
Since the black hole entropy can never be negative, it is tempting to speculate that relations of the type ${V}'_{BH}< c_1 {V}_{BH}$ and ${V}''_{BH} > -c_2 {V}_{BH}$, might hold in certain regions of the moduli space.
Then, using the black hole/flux correspondence we could associate $V_{BH} \leftrightarrow V_{\mathcal{N}=1}$, but we should be careful with signs, not to flip the direction of the bounds. Since $V_{BH}$ is positive definite while $V_{\mathcal{N}=1}$ is negative definite, if we replace $V_{BH} \to - V_{\mathcal{N}=1}$, we would recover the de Sitter conjectures \eqref{ineq-1} as a consequence of positiveness of the black hole entropy. 

We stress that these arguments are at best heuristic and do not provide a derivation for the de Sitter conjectures from black hole physics. At this point, this seems to us a challenging task, whose difficulty is probably related to the fact that such conjectures are not as physically motivated as the distance conjectures that we are going to analyze next.\footnote{Here, we do not mean that there are counter-examples to the de Sitter conjectures, rather that it is unclear at present what is the physical principle behind them, except the lack of explicit de Sitter constructions in string theory. A physical motivation in certain asymptotic regimes has been provided in \cite{Bedroya:2019snp}.}

\subsection{Distance conjectures and a puzzle}\label{distanceconjectures}

We focus now on distance conjectures involving the cosmological constant $\Lambda$, namely the value of $V_{\mathcal{N}=1}$ at the minimum, and the black hole entropy $\mathcal{S}$, namely the value of $V_{BH}$ at the horizon.
For what concerns the cosmological constant, in \cite{Lust:2019zwm} the Anti-de Sitter Distance Conjecture (ADC) was formulated, stating that in the limit $|\Lambda| \to 0$ a tower of states becomes light with mass $m$ scaling as
\begin{equation}
{\rm ADC:}\qquad m\sim |\Lambda|^\alpha\quad{\rm with}\quad\alpha>0\, .
\end{equation}
For what concerns the entropy, the Black Hole Entropy Distance Conjecture (BHEDC) proposed in \cite{Bonnefoy:2019nzv} states that in the limit of large entropy a tower of states becomes light with mass scaling 
as
\begin{equation}
{\rm BHEDC:}\qquad m\sim {\cal S}^{-\beta}\quad{\rm with}\quad\beta>0\, .
\end{equation}
Since the entropy is related to the volume of the moduli space, either by \eqref{SVi} or \eqref{SVii}, the BHEDC is related to limits of large distance on the moduli space \cite{Bonnefoy:2019nzv}.\footnote{In fact, one of the arguments of \cite{Bonnefoy:2019nzv} supporting the BHEDC was precisely to consider the behaviour of the black hole entropy when the volume of the compact manifold, which is a modulus, becomes either big or small. This idea has been further developed in \cite{Hamada:2021yxy} and more recently in \cite{Cribiori:2022cho,Delgado:2022dkz,Cribiori:2022nke}.} 

Both in the case of the cosmological constant, as well as the case of black hole entropy, we assume that the states becoming light are Kaluza-Klein states, with mass\footnote{Restoring the proper units, the Kaluza-Klein mass of the six-dimensional compactification is estimated as $m_{KK} \sim M_s\mathcal{V}^{-\frac16} = M_P g_s\mathcal{V}^{-\frac23}$, where $M_s$ and $g_s=e^\phi$ are the string mass and coupling respectively. However, in these IIA flux vacua, or equivalently at the black hole horizon, the dilaton is stabilzed at $e^\phi = \mathcal{V}^{\frac{1}{2}}$, thus leading to \eqref{mkkV}. When considering different vacua, as in the next section with geometric fluxes, such a simplification might not occur. }
\begin{equation}
\label{mkkV} 
m_{KK}\sim\biggl({\frac{1}{{\cal V}}}\biggr)^\frac16.
\end{equation}
This assumption is supported by the classification of infinite distance limits of \cite{Lee:2019wij}.
Since the cosmological constant and the entropy are the values that the respective scalar potentials assume at a critical point, one could suspect the ADC and the BHEDC could be related. However, when trying to identify these two conjectures, we uncover a discrepancy. 

On the ADC side, we are dealing with flux vacua of the form
\begin{equation}
    AdS_4\times{\cal M}_6 \,.
\end{equation}
In the supersymmetric case, the $AdS_4$ radius, ${\ell^2_{AdS_4}}$, is related to the cosmological constant by
\begin{align}
   -\frac{3}{\ell^2_{AdS_4}}= \Lambda\equiv V_*=-3|{\cal W}|^2 \ . 
\end{align}
In this case, we are interested is in the limit ${\ell^2_{AdS_4}}\to\infty$ in which $\Lambda$ becomes small.

On the BHEDC side, we are dealing with extremal black hole solutions with horizon geometry
\begin{align}
    AdS_2\times S^2 \times {\cal M}_6 \, .
\end{align}
For supersymmetric black holes the radius of $AdS_2$ is equal to radius of $S^2$, namely the horizon radius $r_h$. In terms of the black hole potential, one gets
\begin{align}
    \ell^2_{AdS_2}= r_h^2 = \frac{\cal S}{\pi} = V_{BH}(z_*,\bar z_*) = |{\cal W}|^2 \, ,
\end{align}
i.e.~the black hole entropy scale is set by the scale of the potential $V_{BH}$. 
In this case, we are interested in the limit ${\ell^2_{AdS_2}}\to\infty$ in which ${\cal S}$ and $V_{BH}$  become large.
   
We want to see if the ADC and the BHEDC can be satisfied at the same time for flux vacua and black holes related by the correspondence. For concreteness, we focus on models with cubic prepotentials.  The expressions for $\Lambda$ and ${\cal S}$ in terms of the internal volume, and thus of the Kaluza-Klein scale, are given in \eqref{SVi} and \eqref{SVii}, which are denoted as case (i) and (ii) respectively. For case (i), namely constant electric charges and varying magnetic charges, one has
  \begin{equation}
m_{KK}\sim \left(V_0\right)^{\frac16}q^{-\frac13},
\end{equation}
where we introduced the quantity
\begin{equation}
V_0\sim\sqrt{\frac q6 C_{ijk}p^ip^jp^k}
\end{equation}
encoding the charge dependence of both $|\Lambda|$ and ${\cal S}$. For case (ii), namely constant magnetic charges and varying electric charges, one has 
\begin{equation}
    m_{KK} \sim \left(V_0\right)^{-\frac12} \left(\frac16 C_{ijk} p^ip^jp^k\right)^{\frac{1}{3}}.
    \label{mkkii}
\end{equation}   
These two cases have opposite behaviour for what concerns the conjectures. In case (i), the Kaluza-Klein tower becomes light for small $V_0$, in accordance with the ADC but in disagreement with the BHEDC. In case (ii), the Kaluza-Klein tower becomes light for large $V_0$, in accordance with the BHEDC but in disagreement with the ADC. 
In the following we concentrate on case (ii), namely varying electric charges, in which the large volume limit is given by $q\to \infty$. This is indeed the regime in which the computation of the central charge in the sigma model is under control \cite{Strominger:1996sh,Maldacena:1997de}. 

In such large volume limit, the entropy and the cosmological constant diverge as $V_0 \sim \sqrt{q} \to \infty$. This is unexpected for a large distance limit of $\Lambda$, which should rather approach zero according to the ADC. Instead, here a vanishing cosmological constant seems to be associated to a small flux limit, $q \to 0$, which is not even compatible with flux quantization: fluxes cannot be arbitrarily small.

We can also contrast this behaviour with the well-known case of the $AdS_5\times S^5$ flux vacuum dual to $N$ D3-branes. Here the brane charges (fluxes) are taken to be large, i.e.~$N\to \infty$, in order for the supergravity approximation to be reliable. The $AdS_5$ radius is 
\begin{align}
     \ell^2_{AdS_5} = \sqrt{\alpha'}(4\pi g_{s} N)^{1/4}\gg \sqrt{\alpha'}=\ell_s,
\end{align}
therefore one requires $g_s N\gg 1$ as $N\to\infty$ and $g_s\to 0$. So, this background has indeed large flux, but also small cosmological constant, because
\begin{align}
    |\Lambda| \sim \frac{1}{\sqrt{g_s N}}\, .
\end{align}

Thus, it seems that the ADC and the BHEDC cannot be related via the black hole/flux correspondence. In the next section, we are going to show how to modify such correspondence in order to make the two conjectures compatible with each others.

\subsection{Resolution of the puzzle: the role of geometric fluxes}
\label{geomflux}

In the limit $q \to \infty$ of large volume, the black holes we considered have a consistent behaviour with respect to the BHEDC. In fact, they correspond to BPS solutions with $AdS_2\times S^2 \times {\cal M}_6$ near horizon geometry and electric and magnetic charges $q$ and $p^i$.
On the other hand, the cosmological constant in the effective supergravity theory with $q$ and $p^i$ flux parameters does not possess the expected infinite distance behaviour. 

The reason for this seeming deviation from the ADC is that the superpotential
in \eqref{IIAF} does not lead to a complete stabilization of all moduli. As it is well-known, one can stabilize at most all K\"ahler moduli and one combination of the axio-dilaton and complex structure moduli.\footnote{Indeed, the authors of \cite{DeWolfe:2005uu} considered an orientifold of a toroidal orbifold with rigid complex structure, $h^{2,1}=0$, in order to stabilize all moduli of their IIA setup. This is however a non-generic choice of internal manifold and we would rather like to discuss the general strategy.} Therefore, we are not yet dealing with $AdS_4$ vacua with all moduli fixed, as it is instead assumed in the ADC. The key to solve the discrepancy is thus to extend the supergravity theory by further ingredients allowing for complete moduli stabilization. The black hole/flux correspondence should then take into account these additional contributions.

To stabilize the remaining moduli we introduce geometric fluxes into the superpotential. In the microscopic model, it corresponds to changing the metric of the six-dimensional compact space, moving away from the Calabi-Yau case and introducing curvature. Crucially, these geometric fluxes do not lead to additional electric or magnetic charges in the four-dimensional black holes. Thus, they are not relevant for the black hole attractor mechanism nor they are entering in the black hole central charge $Z$. As we will discuss in the next section, these geometric fluxes corresponds to Kaluza-Klein monopoles in the domain wall picture.

After the inclusion of geometric fluxes, $Z$ and $W$ do not have the same form anymore: $Z(T)$ contains only K\"ahler moduli, whereas $W(T,U,S)$ depends on both K\"ahler and complex structure moduli. Thus, we have to refine the black hole/flux correspondence by identifying the central charge with the complete superpotential and the black hole potential with the complete supergravity potential 
\begin{equation}
\begin{aligned}
    Z(T) \qquad &\Longleftrightarrow \qquad W(T,U,S),\\
    V_{BH}(T) \qquad &\Longleftrightarrow \qquad V_{\mathcal{N}=1}(T,U,S).
\end{aligned}
\end{equation}
This will lead to an agreement between the two distance conjectures we are considering, namely the BHEDC and the ADC.

For concreteness, let us focus again on the simple model with $\mathcal{N}=1$ chiral multiplets $\{T^i, U^i, S\}$, with $i=1,2,3$, already discussed in section \ref{sec_VBHtoVN1} and corresponding to a toriodal type IIA compactification. As explained, the dilaton $S$ and the three complex structure $U^i$ moduli are not yet included in the superpotential \eqref{IIAF}; in the $\mathcal{N}=2$ language they are hyper multiplets. They contribute to the vacuum energy only via the K\"ahler potential and are thus runaway directions of the scalar potential. When turning on geometric fluxes, the model deviates from the flat toroidal compactification and becomes a twisted six-torus with non-trivial (constant) curvature. 

For the purposes of our analysis, it is enough to consider the case in which we identify the moduli as $T^i=T$, $U^i=U$, for $i=1,2,3$. Then, the K\"ahler potential is
\begin{equation}
    K = -3 \log (-i(T-\bar T))-3 \log (-i (U-\bar U))-\log(-i (S-\bar S)) - 3 \log \mathcal{C},
\end{equation}
where, following \cite{Aldazabal:2007sn}, we introduced the flux-dependent normalization $\mathcal{C}$ required for a precise matching with the microscopic model.\footnote{The importance of $\mathcal{C}$ has recently been stressed in relation to scale-separation in \cite{Font:2019uva}.}
The superpotential is instead
\begin{equation}
\label{supogeom}
    \frac{W}{\mathcal{C}} = -q + 3 p T^2 - 3 a ST - 3 b UT,
\end{equation}
and it corresponds to \eqref{IIAF} supplemented by the geometric fluxes $a,b$. These are the structure constants of the twisted six-torus with vielbeins 1-forms $\eta^m$, $m=1,\dots, 6$, namely they enter the relation $d\eta^1 = -a \eta^5 \wedge \eta^6- b \eta^2 \wedge \eta^3$.
The internal volume is given by \cite{Aldazabal:2007sn}
\begin{align}\label{vol-C}
    {\cal V}&=\mathcal{C}({\rm Im}T)^3, \qquad \mathcal{C}=\frac{(4\pi)^4}{(4ab^3)^{\frac 32}},
\end{align}

The model has a supersymmetric anti-de Sitter vacuum in which axions are stabilized at
\begin{equation}
    {\rm Re} \,S = {\rm Re}\, T= {\rm Re}\, U=0
\end{equation}
(in fact there is a generalization of this class where $a \,{\rm Re}\, S+b\, {\rm Re} \,U=0$) and the saxions are given by
\begin{equation}
   s\equiv {\rm Im}\, S=\frac{2}{3a}\sqrt{qp}, \qquad t\equiv {\rm Im }\,T =\frac 13 \sqrt{\frac{q}{p}}, \qquad u \equiv {\rm Im }\, U=\frac 2b \sqrt{qp}.
\end{equation}
The cosmological constant is
\begin{equation}
    \Lambda =-\frac{27 ab^3}{128{\cal C}\sqrt{q^3p}}
\end{equation}
and, for fixed $p$ and metric fluxes, it scales like 
\begin{align}
\Lambda&\sim {q^{-\frac32}} \ .
\end{align}  
Therefore, the cosmological constant becomes now smaller when the flux charge $q$ is increased, unlike the behaviour in \eqref{mkkii}.
On the vacuum, the volume of the internal manifold is
\begin{equation}
    \mathcal{V}_* = \frac{\mathcal{C}}{27}\left(\frac qp\right)^{\frac 32}
\end{equation}
and, for fixed $p$ and metric fluxes, it scales as
\begin{align}
    {\cal V}_* &\sim q^{3/2} \ .
\end{align}

Summarising, we found that a fixed $p$ and $\mathcal{C}$ we have 
\begin{align}
       q\to\infty \qquad \Rightarrow \qquad {\cal V}_*\to\infty \quad \& \quad \Lambda \to 0 \ ,
\end{align}
thus solving the puzzle one would get by simply considering the model without the fields $U^i$ and $S$ in the superpotential.
Furthermore, taking into account the string coupling constant, one can even concretely show that the geometric flux dependence drops out from the ratio between the Kaluza-Klein mass and the cosmological constant \cite{Font:2019uva}, namely
\begin{equation}
m_{KK}\sim |\Lambda|^{\frac 12}\, ,
\end{equation}
in precise agreement with the strong version of the ADC.

\subsection{Including the domain walls}
\label{KKdw}

In order to complete our analysis of the correspondence between black holes, domain walls and fluxes, we have to explain how the domain walls sourcing the geometric fluxes $a$ and $b$ are constructed. As we will see, they corresponds to Kaluza-Klein monopoles in the microscopic model. 

After the introduction of geometric fluxes, the internal manifold becomes a twisted torus and it can be described by a basis of vielbein 1-forms $\eta^m$, $m=1,\dots,6$, satisfying the relations
\begin{equation}
d\eta^m=-{\frac12} \omega_{np}^m   \, \eta^n\wedge \eta^p\, .
\end{equation}
The quantities $\omega^m_{np}$ are quantized structure constants to be identified with $a$ and $b$ in our concrete example. By reviewing their origin from the T-dual type IIB setup, we can explain how to construct the domain walls we are looking for. We follow mainly \cite{Camara:2005dc,Aldazabal:2006up}.

First, let us consider the $a$-fluxes. They are given by three structure constants
\begin{equation}
\left(
    \begin{array}{c}
    \omega^1_{56}\\
    \omega^2_{64}\\
    \omega^3_{45}
    \end{array}
    \right)=
    \left(
    \begin{array}{c}
    a_1\\
    a_2\\
    a_3
    \end{array}
    \right),
\end{equation}
upon identifying $a_i \equiv a$. Their presence result in a twist of the toriodal metric of the form
\begin{equation}
\begin{aligned}
ds^2 &=   (dy^1+a_1y^6dy^5)^2+(dy^2+a_2y^4dy^6)^2+(dy^3+a_3y^5dy^4)^2 \\
&+ (dy^4)^2  + (dy^5)^2  + (dy^6)^2.  
\end{aligned}
\end{equation}
This can be reproduced by following the usual Buscher rules starting from a type IIB setup with a flat toroidal metric but non-trivial H-flux
\begin{equation}
H_3=-a_1 dy^1\wedge dy^5\wedge dy^6-a_2 dy^4\wedge dy^2\wedge dy^6-a_3 dy^4\wedge dy^5\wedge dy^3\, .
\end{equation}
Thus, in type IIB the $a$-fluxes correspond to three specific components of $H_3$ which are sourced by three different NS5-branes.
\vspace{16pt}
\begin{center}
				\begin{tabular}{  l | *{4}{c} | *{6}{c}  }
				  \toprule
				  Domain Wall  & $x^0$ & $x^1$ & $x^2$ & $x^3$ & $y^1$ & $y^2$ & $y^3$ & $y^4$ & $y^5$ & $y^6$  \\ \midrule
				  NS$_1$ & X & $\cdot$ & X & X& $\cdot$  & X &  X & X & $\cdot$ & $\cdot$ \\ \midrule
				  NS$_2$ & X & $\cdot$ & X & X & X & $\cdot$ & X &  $\cdot$ &  X & $\cdot$ \\ \midrule
				  NS$_3$ & X & $\cdot$ & X & X & X &  X  & $\cdot$ & $\cdot$ &  $\cdot$ & X \\  
				  \bottomrule
				\end{tabular}
			  \end{center}
			\vskip0.4cm
   \vspace{16pt}
Since they wrap three different 3-cycles of the internal geometry, they are in fact domain walls in four space-time dimensions. Going back to type IIA, the NS5-branes become three different Kaluza-Klein monopoles.

\vspace{15pt}

\begin{center}
				\begin{tabular}{  l | *{4}{c} | *{6}{c}  }
				  \toprule
				  Domain Wall  & $x^0$ & $x^1$ & $x^2$ & $x^3$ & $y^1$ & $y^2$ & $y^3$ & $y^4$ & $y^5$ & $y^6$  \\ \midrule
				  KK$_1$ & X & $\cdot$ & X & X& $\bullet$  & X &  X & X & $\cdot$ & $\cdot$ \\ \midrule
				  KK$_2$ & X & $\cdot$ & X & X & X & $\bullet$ & X &  $\cdot$ &  X & $\cdot$ \\ \midrule
				  KK$_3$ & X & $\cdot$ & X & X & X &  X  & $\bullet$ & $\cdot$ &  $\cdot$ & X \\  
				  \bottomrule
				\end{tabular}
			  \end{center}
			\vskip0.4cm
Here, $\bullet$ denotes the so called NUT direction, namely the direction of the monopoles isometry. Thus, we see that the domain walls sourcing the three geometric fluxes $a^i$ are given by Kaluza-Klein monopoles in the microscopic models. In a generic setup, these can be included in addition to the D-brane domain walls already discussed in section  \ref{domainwalls}.

Next, let us look at the $b$-fluxes. They are given by nine structure constants
\begin{equation}
\left(
\begin{array}{ccc}
-\omega_{23}^1 & \omega_{53}^4 & \omega_{26}^4\\
\omega_{34}^5 & -\omega_{31}^2 & \omega_{61}^5\\
\omega_{42}^6 & \omega_{15}^6 & -\omega_{12}^3\\
\end{array}
\right)=
\left(
\begin{array}{ccc}
b_{11} & b_{12} & b_{13}\\
b_{21} & b_{22} & b_{23}\\
b_{31} & b_{32} & b_{33}\\
\end{array}
\right)\equiv
\left(
\begin{array}{ccc}
-b & b & b\\
b & -b & b\\
b & b & -b\\
\end{array}
\right),
\end{equation}
upon identifying $b_{ii}=-b$ and $b_{ij}=b$ for $i\neq j$. Similarly to what discussed before, in the T-dual type IIB setup with flat toroidal metric they originate from four non-vanishing components of $H_3$
\begin{equation}
\begin{aligned}
H_3=&-(b_{11}+b_{22}+b_{33}) dy^1\wedge dy^2\wedge dy^3  \\
&+(b_{12}-b_{21})dy^3 \wedge dy^4 \wedge dy^5 \\
&+ (b_{31} - b_{13}) dy^2 \wedge dy^4 \wedge dy^6 \\
&+ (b_{23}-b_{32}) dy^1\wedge dy^5 \wedge dy^6.       
\end{aligned}
\end{equation}
These components are sourced by NS5-branes collected in the following table.
\begin{center}
				\begin{tabular}{  l | *{4}{c} | *{6}{c}  }
				  \toprule
				  Domain Wall  & $x^0$ & $x^1$ & $x^2$ & $x^3$ & $y^1$ & $y^2$ & $y^3$ & $y^4$ & $y^5$ & $y^6$  \\ \midrule
				  NS$_{11}$ & X & $\cdot$ & X & X & $\cdot$  & $\cdot$ &  $\cdot$ & X & X & X \\ \midrule
                NS$_{22}$ & X & $\cdot$ & X & X & $\cdot$  & $\cdot$ &  $\cdot$ & X & X & X \\ \midrule
                NS$_{33}$ & X & $\cdot$ & X & X & $\cdot$  & $\cdot$ &  $\cdot$ & X & X & X \\ \midrule
				  NS$_{12}$ & X & $\cdot$ & X & X & X & X & $\cdot$ &  $\cdot$ &  $\cdot$ & X\\ \midrule
                NS$_{21}$ & X & $\cdot$ & X & X & X & X & $\cdot$ &  $\cdot$ &  $\cdot$ & X\\ \midrule
                NS$_{31}$ & X & $\cdot$ & X & X & X &  $\cdot$  & X & $\cdot$ &  X & $\cdot$ \\ \midrule
				  NS$_{13}$ & X & $\cdot$ & X & X & X &  $\cdot$  & X & $\cdot$ &  X & $\cdot$ \\ \midrule
				  NS$_{23}$ & X & $\cdot$ & X & X & $\cdot$ & X  & X & X &  $\cdot$ & $\cdot$ \\  \midrule
				  NS$_{32}$ & X & $\cdot$ & X & X & $\cdot$ & X  & X & X &  $\cdot$ & $\cdot$ \\  
				  \bottomrule
				\end{tabular}
			  \end{center}
			\vskip0.4cm
Here, we used the notation NS$_{ij}$ to refer to brane associated to $b_{ij}$. Notice that ${\rm NS}_{11}={\rm NS}_{22}={\rm NS}_{33}$, ${\rm NS}_{12}={\rm NS}_{21}$, ${\rm NS}_{13}={\rm NS}_{31}$,  ${\rm NS}_{23}={\rm NS}_{32}$ as brane configurations, thus only four out of nine NS5-branes are really different. However, when T-dualizing back to type IIA we can produce nine different Kaluza-Klein monopoles depending on the NUT directions. We report them below, following a similar notation. These are again domain walls in the four-dimensional spacetime and can be added to the general discussion of section \ref{domainwalls}. 
\begin{center}
				\begin{tabular}{  l | *{4}{c} | *{6}{c}  }
				  \toprule
				  Domain Wall  & $x^0$ & $x^1$ & $x^2$ & $x^3$ & $y^1$ & $y^2$ & $y^3$ & $y^4$ & $y^5$ & $y^6$  \\ \midrule
				  KK$_{11}$ & X & $\cdot$ & X & X &  $\bullet$ & $\cdot$ &  $\cdot$ & X & X & X \\ \midrule
                KK$_{22}$ & X & $\cdot$ & X & X & $\cdot$  &  $\bullet$ &  $\cdot$ & X & X & X \\ \midrule
                KK$_{33}$ & X & $\cdot$ & X & X & $\cdot$  & $\cdot$ &   $\bullet$ & X & X & X \\ \midrule
				  KK$_{12}$ & X & $\cdot$ & X & X & X & X & $\cdot$ &  $\bullet$ &  $\cdot$ & X\\ \midrule
                KK$_{21}$ & X & $\cdot$ & X & X & X & X & $\cdot$ &  $\cdot$ &  $\bullet$ & X\\ \midrule
                KK$_{31}$ & X & $\cdot$ & X & X & X &  $\cdot$  & X & $\cdot$ &  X &  $\bullet$ \\ \midrule
				  KK$_{13}$ & X & $\cdot$ & X & X & X &  $\cdot$  & X & $\bullet$ &  X & $\cdot$ \\ \midrule
				  KK$_{23}$ & X & $\cdot$ & X & X & $\cdot$ & X  & X & X &   $\bullet$ & $\cdot$ \\  \midrule
				  KK$_{32}$ & X & $\cdot$ & X & X & $\cdot$ & X  & X & X &  $\cdot$ &  $\bullet$ \\ 
				  \bottomrule
				\end{tabular}
			  \end{center}
			\vskip0.4cm

\section{Asymptotically AdS black holes and the ADC}
\label{adcaadsbh}

Finally, we would like to comment on possible extensions of our considerations to BPS black holes that are asymptotic to anti-de Sitter spacetime in four-dimensional $\mathcal{N}=2$ supergravity. 
A detailed investigation of such models within the Swampland Program is left for the future. 
These solutions are known since the works of \cite{Cacciatori:2009iz,DallAgata:2010ejj,Hristov:2010ri} and their microscopic description has been obtained holographically from the dual ABJM theory in \cite{Benini:2015eyy,Benini:2016rke}.

In the simplest case, it is enough to consider supergravity with an R-symmetry gauging leading to an anti-de Sitter vacuum. In this theory, the black hole is an interpolating geometry between an asymptotic $AdS_4$ region and a near horizon $AdS_2\times S^2$ region. In this sense, we can understand them as domain walls between two anti-de Sitter spaces of different dimension. Such an interpretation is analogous to BPS black holes in asymptotically flat space, which are supersymmetric solitons interpolating between maximally symmetric geometries at asymptotic infinity and at the horizon \cite{Ferrara:1995ih}, and it is also consistent with the dual holographic interpretation of these states as RG-flows across dimensions. 

At infinity, the scalar fields take the values set by the anti-de Sitter extremum condition \eqref{AdS4-extr-cond}, which is now conveniently re-expressed as
\begin{align}\label{extr-L}
\partial_{i}|\mathcal{L}|=0 \ .
\end{align}
Here, we denoted $\mathcal{L}=g_\Lambda L^\Lambda(z^i, \bar z^{\bar \imath})-\tilde{g}^\Lambda F_\Lambda(z^i, \bar z^{\bar \imath})$, which is precisely the superpotential $\mathcal{W}$ in \eqref{Wgauged}. The different notation is now helpful in avoiding confusion between black hole charges $q_\Lambda,p^{\Lambda}$, namely U$(1)$ fluxes in space-time,  and gauging parameters $g_\Lambda,\tilde{g}^\Lambda$,  namely U$(1)$ fluxes in the internal dimensions. Notice that both appear in the Lagrangian of $\mathcal{N}=2$ U$(1)_R$ gauged supergravity, the former as conserved abelian charges in four dimensions, the latter as gauging derived from internal fluxes.

From \eqref{extr-L}, one recognizes that $\mathcal{L}$ plays the same role as the central charge ${Z}$ of BPS black holes in asymptotically flat space-time, for it fixes the scalar fields at infinity solely in terms of the gauging parameters $g_\Lambda,\tilde{g}^\Lambda$. 
The algebraic content of this  equation is that, if the gauging parameters $g_\Lambda,\tilde{g}^\Lambda$ are related to quantized fluxes of the higher dimensional theory, then the cosmological constant
\begin{align}
\Lambda = -3 |\mathcal{L}|^2 \ ,
\end{align}
is also quantized. 

In order for the black hole to have non-vanishing entropy, supersymmetry introduces a constraint involving $p^\Lambda$ and $g_\Lambda$. For a spherical black hole horizon (and in a specific symplectic frame), this is \cite{DallAgata:2010ejj}\footnote{We are only considering black holes with a spherical horizon. In anti-de Sitter one can also have black holes with planar horizon, but they cannot be connected to BPS black holes in Minkowski, since these can only have spherical horizon. In fact, \eqref{constr-susy-BH-AdS} is a special case of $p^\Lambda g_\Lambda = -k$, where $k$ is a parameter related to the curvature of the factor multiplying $AdS_2$ in the horizon geometry. More in general, in the non-supersymmetric case one encounters the quantization condition $p^\Lambda g_\Lambda = n \in \mathbb{Z}$, from which conclusions similar to those below might be drawn \cite{Klemm:2012vm,Gnecchi:2012kb}. }
\begin{align}\label{constr-susy-BH-AdS}
    p^\Lambda g_\Lambda= -1 \ .
\end{align}
This constraint has a nice interpretation within the Swampland Program, for it prevents the existence of a smooth interpolation between asymptotically anti-de Sitter and asymptotically Minkowksi BPS black holes, which would correspond to $g_\Lambda= 0$.\footnote{Recall that in gauged supergravity the scalar potential is always quadratic in the gauge coupling. Thus, $g_\Lambda=0$ gives a Minkowski vacuum.} Indeed, the limit $g_\Lambda \to 0$, which is the relevant one for the ADC, must be accompanied by $p^\Lambda \to \infty$ in order for \eqref{constr-susy-BH-AdS} to be satisfied. 

Naively, one might expect the large flux limit to be problematic from an effective field theory point of view. In fact, we saw that it corresponds to the large entropy limit of the asymptotically flat black hole configuration which leads to de-compactification.
We can make such an expectation precise by looking at the attractor equations for these asymptotically anti-de Sitter black hole solutions.
They are given by \cite{DallAgata:2010ejj}
\begin{align}
    \partial_i|Z| = R_{S^2}^2\ \partial_i|\mathcal{L}| \ , \qquad
    R_{S^2}^2 =  \frac{|Z|}{|\mathcal{L}|}   \ ,
\end{align}
where $R_{S^2}$ is the radius of the $S^2$ at the horizon. Let us consider for concreteness the cubic model with only three fields that are identified, $X^1=X^2=X^3$, and with the charges $g_1=g_2=g_3\equiv g$ and $p^1=p^2=p^3\equiv p$, besides $g_0$ and $p^0$. The black hole entropy is given explicitly by \cite{Gnecchi:2012kb}
\begin{align}
    \frac{\mathcal{S}}{\pi} = R_S^2 = \frac{3^{3/2}}{8 \sqrt{g_0g^3}}\left( \sqrt{1+4 pg}+3 \sqrt{1+\frac43 g p}  \right)^{\frac12}
    \left( \sqrt{1+4 pg}- \sqrt{1+\frac43 gp}  \right)^{\frac32}.
\end{align}
We can see that the limit of small anti-de Sitter cosmological constant, namely $g_{\Lambda} \to 0$ with $p^\Lambda g_\Lambda =-1$, corresponds to a de-compactification limit of the horizon $S^2$ which signals the breakdown of the effective field theory description according to the BHEDC.\footnote{A sickness in the interpolation between anti-de Sitter and flat space is expected already from the the supergravity construction. In fact, the Minkowski BPS black hole is a 1/2-BPS solution (with a horizon enhancing supersymmetry to full $\mathcal{N}=2$), while the $AdS_4$ black holes are 1/4-BPS solutions and the horizon is a 1/2-BPS geometry of the gauged theory when magnetic fluxes are turned on. We are suggesting here that the constraint that supersymmetry imposes on the parameters already takes care of the obstruction in parameter space of the solution.} Notice that the limit has to be taken with care, namely $g_0,g \to 0$ and $p^0, p \to \infty$ in such a way that $g p \sim \text{const}\neq 0$.
In this limit, we expect an infinite tower of states to become light.
While the investigation of the precise nature of these states is left for future work, here we want to stress the connection between the appearance of a constraint \eqref{constr-susy-BH-AdS} and the ADC. 
In fact, \eqref{constr-susy-BH-AdS} is a necessary condition for the existence of these BPS black holes, that imposes a topological twist in the dual SCFT. Examples of these kind of domain walls across dimensions have appeared first in five dimensions as black strings in $AdS_5$, in the context of $c$-extremization \cite{Benini:2012cz,Benini:2013cda}, and they also realize the ADC via an analogous flux quantization condition. 

There are, however, supersymmetric solutions in four dimensions with an $AdS_2$ factor that preserve supersymmetry without a topological twist. Examples are supersymmetric, four dimensional black holes in anti-de Sitter with a well defined Minkowski limit, but they require a nonzero rotation parameter, which also goes to zero in the flat space limit. 
Another example is the class of solutions recently obtained in the context of holography, which are characterized by their horizon geometry being $AdS_2\times \Sigma$ where $\Sigma$ is a \textit{spindle} \cite{Ferrero:2020laf}. 
In general, spindle horizons are topologically a sphere, with orbifold (conical) singularities at the North and South poles (parametrized by two co-prime integers $n_1,n_2$). They uplift to regular geometries in ten or eleven dimensions, corresponding to wrapped D3, M2, D4 or M5 branes \cite{Ferrero:2020laf,Ferrero:2021wvk,Couzens:2021rlk,Boido:2021szx}. 
They are rotating geometries that can preserve supersymmetry in two ways \cite{Ferrero:2021etw}. One way is a generalization of the topological twist, that require the Killing spinors to be invariant under the $U(1)$ isometry of the rotational symmetry of the spindle. 
In this case, in the limit $n_1=n_2=1$ one recovers the spherical horizon solutions of the $AdS_4$ black holes described in this section, which exist both with angular momentum as well as a static solution. The ADC is realized again by a constraint among the charges like in \eqref{constr-susy-BH-AdS}, and in the case of the spindle it can be generalized as
\begin{align}
    p_\Lambda g^\Lambda &= - (n_1+n_2)  \ .
\end{align}
The spherical case corresponds to the limit $n_1=n_2=1$.
The other possibility is that supersymmetry is realized via a so-called \textit{anti-twist}, and in this case the Killing spinors transform non-trivially with respect to the spindle isometries. This case also requires a constraint among the charges, which can be written as
\begin{align}
    p^\Lambda g_\Lambda  = n_1-n_2 \ .
\end{align}
Notice that if we had $n_1=n_2$, there could be room for taking $g_\Lambda\to0$ at fixed magnetic charges. However, this case differs from our discussion since, in general, spindle solutions are rotating, and it is known that rotating black holes solutions in anti-de Sitter admit a flat space limit, even in the case of spherical horizons. We are not interested in the case where rotation cannot be turned off as an independent parameter from the cosmological constant, 
so we leave this case out of our discussion and to a future investigation. 

\section{Conclusions and future directions}
\label{sec:conclusions}

In this work, we reviewed and extended the correspondence between BPS black holes, domain walls and string flux vacua, according to which, modulo proper identifications and adjustments, the same effective supergravity action can be conveniently employed to describe three different microscopic systems realized in string theory by specific D-brane configurations.
In this framework, we investigated and related to one another swampland conjectures, especially those involving a potential for the scalar fields. 
The main motivation behind our analysis is to understand if and to what extent swampland conjectures can be derived or at least motivated by properties of black hole physics, such as finiteness of the entropy. 

Concretely, we focussed on type IIA compactifications, where we looked at an explicit toroidal orientifold, and worked out the dictionary between black holes, domain walls and flux vacua in detail. The three systems are described by a function $\mathcal{W}$, which can be identified in turn with the black hole central charge $Z_{BH}$, the domain wall tension $Z_{DW}$, or the flux potential, $W_{flux}$. Such function depends on the scalar fields but also on certain parameters, which are respectively  the black hole or domain wall charges, or the supergravity gauge couplings. Guided by $\mathcal{W}$, we proposed precise identifications between other quantities on the different sides of the correspondence, such as the black hole potential or entropy, and the supergravity potential or (gravitino) gauge coupling. We thus built a dictionary, which can be used to test or motivate swampland conjectures from black hole physics.  

As an application, we looked at conjectures involving a scalar potential. While for the de Sitter conjectures we have been able only to propose some speculations, for the ADC and the BHEDC our analisys has been quantitative. We noticed that when comparing these conjectures via the correspondence between black holes and flux vacua one encounters a mismatch, for the entropy and the cosmological constant do not share the same behaviour when the charges are sent to infinity (corresponding to infinite distance in the moduli space \cite{Bonnefoy:2019nzv}). We identified the reason for this discrepancy in the fact that, on the supergravity side, some of the moduli are still not stabilized and thus the ADC cannot be straightforwardly applied. Therefore, to achieve full moduli stabilization, we extended the correspondence in order to include geometric fluxes and the corresponding Kaluza-Klein monopole domain walls sourced by them. After this step, we could verify that the cosmological constant and the entropy finally had the same behaviour for large charges (or moduli), supporting the fact that the ADC and the BHEDC are closely related.

While for the main part of the work we considered asymptotically flat black holes, namely black holes in ungauged supergravity, in the last section we looked at asymptotically anti-de Sitter black holes, namely black holes in gauged supergravity. These have an $AdS_2\times S^2$ horizon, which is however different from that of their asymptotically flat counterparts, and, crucially, they are subject to a constraint involving black hole charges and supergravity gauge couplings. Interestingly, the form of this constraint is such that it forbids a smooth interpolation between the asymptotic $AdS_4$ and Minkowski spacetimes, for the latter would correspond to a black hole with infinite entropy. We interpreted the presence of this constraint and its consequences as an intrinsic realization of the ADC in these setups.

It would be interesting to understand if the correspondence between flux vacua and BPS black holes here investigated can be extended to include physical processes connecting $AdS_4$ vacua to $AdS_2 \times S^2$ backgrounds of ungauged supergravity. For example, one such process could be given by brane-flux annihilation. In such case, a thin bubble, corresponding to a spherical domain wall, should exist in the effective theory as solution of the (ten-dimensional) field equations describing a dynamical transition between a black hole geometry and a corresponding flux vacuum. This could be detected by branes nucleation in higher dimensions. Notice that inside the bubble the gauge fields would be non-zero, while outside of the bubble they would vanish identically. 
However, these processes would describe an instability which should be absent for supersymmetric vacua, like in this case. We leave a systematic investigation of this and other possibilities for future work.

\paragraph{Acknowledgments.}
We would like to thank Markus Dierigl for discussions, collaboration at the initial stages of the project and for comments on the manuscript. We would like to thank Davide Cassani for interesting discussions. 
A.G.~would like to thank the Max Planck Institute for Physics, Munich, for providing an exciting and inspiring atmosphere during her research work there.
The work of N.C.~is supported by the Alexander-von-Humboldt foundation.
The work of D.L.~is supported by the Origins Excellence Cluster and by the German-Israel-Project (DIP) on Holography and the Swampland.

\appendix

\section{Special geometry of cubic prepotentials}
\label{app_skgeom}

In this appendix we collect more details on the class of models with cubic prepotential which is used as concrete example in the main text.

The classical prepotential arising from type IIA string theory compactified on a Calabi-Yau manifold is
\begin{equation}
    F(X)=-\frac16 C_{ijk} \frac{X^i X^j X^k}{X^0} = -\frac16 C_{ijk} z^i z^j z^k\, (X^0)^2.
\end{equation}
The symplectic sections are constructed out of $X^\Lambda=(X^0,X^i)$ and
\begin{align}
    F_\Lambda= X^0\left(\frac16 C_{ijk}z^i z^j z^k,-\frac12 C_{ijk} z^jz^k\right),
\end{align}
while the physical scalars in the vector multiplets are given by $z^i = \frac{X^i}{X^0}$. 

The K\"ahler potential is
\begin{equation}
\begin{aligned}
    K &= -\log\left[i\left(\bar X^\Lambda F_\Lambda - X^\Lambda \bar F_\Lambda\right)\right]\\
    &=-\log\left[\frac i6 C_{ijk}(z^i-\bar z^i)(z^j-\bar z^j)(z^k-\bar z^k) X^0\bar X^0\right]\\
    &=-\log\left[\frac43 C_{ijk}t^it^jt^k \, X^0\bar X^0\right]\\
    &=-\log\left[8\mathcal{V}X^0 \bar X^0\right],
\end{aligned}
\end{equation}
where we split the complex scalars into real and imaginary parts, 
\begin{equation}
    z^i = b^i + i t^i,
\end{equation}
and
\begin{equation}
\mathcal{V} = \frac16 \int_{CY} J \wedge J \wedge J = \frac16 C_{ijk}t^it^jt^k
\end{equation}
is the volume of the Calabi-Yau with K\"ahler 2-form $J$.\footnote{The coordinates $t^i$ are relates to the very special geometry coordinates $\hat t^i$ of five-dimensional supergravity by $\hat t^i={\cal V}^{-1/3}t^i$; they satisfy $\frac{1}{6}C_{ijk}\hat t^i\hat t^j\hat t^k=1 \ .$}
One can calculate
\begin{align}
K_i &=\partial_iK= -\frac i2 e^K C_{ijk}    (z^j-\bar z^j)(z^k-\bar z^k) \, X^0\bar X^0= 2i e^KC_{ijk} t^j t^k \, X^0\bar X^0,\\
K_{\bar\jmath} &=\partial_{\bar\jmath}K= \frac i2 e^K C_{ijk}    (z^j-\bar z^j)(z^k-\bar z^k) \, X^0\bar X^0= -2i  e^K C_{ijk} t^j t^k\, X^0\bar X^0,
\end{align}
and then the K\"ahler metric is explicitly
\begin{align}
g_{i\bar \jmath} &= K_i  K_{\bar \jmath} + i e^K C_{ijk} (z^k-\bar z^k)\, X^0\bar X^0.
\end{align}

It can be convenient to pass to the real basis of fields $t^i$, since their axionic partners $b^i$ do not enter the K\"ahler potential. To this purpose, we define the symmetric matrix
\begin{equation}
     n_{ij} =i e^K C_{ijk} (z^k-\bar z^k)\, X^0\bar X^0= -2 e^K C_{ijk}t^k X^0\bar X^0
\end{equation}
and we have the relations
\begin{align}
    t_i&=-\frac12 n_{ij}t^j,\\
    t^i &= -2 n^{ij}t_j,\\
    t^it_i &= \frac34,\\
    t^i n_{ij}t^j &= -\frac32,\\
     t_i n^{ij}t_j &= -\frac38,
\end{align}
where $K_i = -i n_{ij}t^j = 2i t_i$ and $n_{ij}n^{jk}=\delta^i_k$.
In the new basis, the K\"ahler metric and its inverse are thus
\begin{align}
g_{i \bar \jmath} &= 4 t_i t_j + n_{ij},\\
g^{i \bar \jmath} &= 2 t^i t^j + n^{ij},
\end{align}
and one can check that
\begin{equation}
K_i g^{i\bar \jmath}K_{\bar \jmath} = 3.
\end{equation}
Other useful relations are
\begin{align}
    t^k\frac{\partial}{\partial t^k} t_i = - t_i,\qquad  t^k\frac{\partial}{\partial t^k} n_{ij} = - 2n_{ij},
\end{align}
leading to
\begin{equation}
\label{tdg}
    t^k\frac{\partial}{\partial t^k} g_{i\bar \jmath} = -2 g_{i\bar\jmath}.
\end{equation}

We are interested in BPS black holes with non-vanishing charges $-q_0=q>0$ and $p^i>0$. The generalization of the formulae below to the full set of non-vanishing charges can be found in \cite{Ceresole:2007rq}, with $t^i_\text{here} = -t^i_\text{there}$.
The $\mathcal{N}=2$ central charge in our case is 
\begin{equation}
\begin{aligned}
Z &= q_\Lambda L^\Lambda - p^\Lambda M_\Lambda \\
&= X^0 e^\frac{K}{2} \left(-q + \frac12 C_{ijk}z^i z^j p^k\right)\\
&=X^0 e^\frac{K}{2} \left(-q + \frac12 C_{ijk}(b^ib^j-t^it^j) p^k + iC_{ijk} b^i t^j p^k\right)
\end{aligned}
\end{equation}
and its covariant derivative is
\begin{align}
    D_i Z = \left(\partial_i+\frac12 \partial_iK\right)Z = X^0 e^\frac{K}{2}\,C_{ijk} z^j p^k + K_i Z.
\end{align}
Furthermore, one can calculate
\begin{equation}
\begin{aligned}
Z \bar Z = |X^0|^2 e^K\left[ q^2-q C_{ijk}(b^ib^k-t^it^j)p^k +\left(\frac12 C_{ijk}(b^ib^j-t^it^j)p^k\right)^2 + \left(C_{ijk}b^it^jp^k\right)^2\right],
\end{aligned}
\end{equation}
which does not contain linear terms in the axions $b^i$. 
Setting $b^i=0$, these quantities simplify to
\begin{align}
Z|_{b^i=0} &=- X^0e^\frac{K}{2} \left(q+2 \mathcal{V}(t^in_{ij}p^j)\right),\\
Z \bar Z|_{b^i=0} &=\frac{q^2}{8\mathcal{V}} - {\frac12} q\,(t^in_{ij}p^j) + \frac12 \mathcal{V}\,(p^in_{ij}t^j)^2.
\end{align}
Similarly, one can calculate
\begin{equation}
\begin{aligned}
|D_i Z|^2 &= g^{i\bar \jmath}   D_i Z \bar D_{\bar \jmath} \bar Z\\
&=X \bar X^0 e^K g^{i \bar \jmath} C_{imn}z^m p^n C_{jrs}\bar z^r p^s + \left(2i\bar X^0 e^\frac{K}{2} Z \, t^i C_{ijk}\bar z^j p^k+c.c.\right) + 3Z \bar Z\\
&= X \bar X^0 e^K \bigg[g^{i \bar \jmath} C_{imn}C_{jrs} (b^mb^r+t^mt^r)p^np^s\\
&\quad +2 C_{ijk}t^ip^j \left(-2qt^k+C_{mnr}p^r\left(b^kb^mb^n+b^kt^mb^n+t^kb^mb^n-t^kt^mt^n\right)\right) \\
&\quad+ 3\left( q^2-q C_{ijk}(b^ib^j-t^it^j)p^k +\left(\frac12 C_{ijk}(b^ib^j-t^it^j)p^k\right)^2 + \left(C_{ijk}b^it^jp^k\right)^2\right)\bigg]
\end{aligned}
\end{equation}
and notice that it does not contain linear terms in the axions $b^i$ either. Setting $b^i=0$, it simplifies to
\begin{equation}
    \begin{aligned}
    |D_iZ|^2|_{b^i=0}&=\frac38 \frac{q^2}{\mathcal{V}}+\frac12 q (p^in_{ij}t^j) + 2\mathcal{V}(p^in_{ij}p^j)+\frac32 \mathcal{V}(p^in_{ij}t^j)^2.
    \end{aligned}
\end{equation}
The black hole potential is
\begin{equation}
    V_{BH} = |D_iZ|^2 + |Z|^2
\end{equation}
and it does not contain any term linear in the axions $b^i$ since neither of the two terms does. Its form at $b^i=0$ is
\begin{equation}
\begin{aligned}
    V_{BH}|_{b^i=0}&=\frac12 \frac{q^2}{\mathcal{V}}+2 \mathcal{V}(p^in_{ij}t^j)^2+2\mathcal{V}(p^in_{ij}p^j)\\
    &=\frac12 \frac{q^2}{\mathcal{V}}+2\mathcal{V}(p^ig_{i\bar\jmath}p^{\bar \jmath}).
\end{aligned}
\end{equation}

Finally, we can calculate
\begin{equation}
    V_{\mathcal{N}=1} = |DZ|^2-3 |Z|^2 = V_{BH} - 4 |Z|^2,
\end{equation}
which again does not contain linear terms in the axions $b^i$. Setting $b^i=0$, it simplifies to
\begin{equation}
\begin{aligned}
     V_{\mathcal{N}=1}|_{b^i=0} &= 2\mathcal{V} (p^i n_{ij} p^j) + 2q (p^i n_{ij}t^j)\\
     &=-\frac12 C_{ijk}p^ip^j t^k -\frac{q}{2\mathcal{V}}C_{ijk}t^it^j p^k.
\end{aligned}
\end{equation}

\bibliographystyle{JHEP}
\bibliography{papers.bib}

\end{document}